\documentclass[12pt,preprint]{emulateapj}
\usepackage{amsfonts}
\usepackage{amssymb}
\usepackage{amsmath}

\shorttitle{NGC7538 - IRS1: Outflow and Protostars
in NGC 7538 IRS 1}
\shortauthors{Zhu et al.}

\begin{document}

\title{Sub-arcsec Observations of NGC\,7538 IRS\,1: Continuum Distribution
and Dynamics of Molecular Gas}
\author{Lei Zhu\altaffilmark{1,2,3},
Jun-Hui Zhao\altaffilmark{2},
M. C. H. Wright\altaffilmark{4},
G\"oran Sandell\altaffilmark{5},
Hui Shi\altaffilmark{1,6},
Yue-Fang Wu\altaffilmark{3},
Crystal Brogan\altaffilmark{7} \&
Stuartt Corder\altaffilmark{7}}
\altaffiltext{1}{National Astronomical Observatories, Chinese
Academy of Sciences, A20 Datun Road, Chaoyang District, Beijing
100012, China; lzhu@nao.cas.cn}
\altaffiltext{2}{Harvard-Smithsonian Center for Astrophysics, 60
Garden Street, Cambridge, MA 02138, USA}
\altaffiltext{3}{Department of Astronomy, Peking University, Beijing
100871, China}
\altaffiltext{4}{Department of Astronomy, University of California,
Berkeley, Berkeley, CA 94720, USA}
\altaffiltext{5}{SOFIA-USRA, NASA Ames Research Center,  MS
232-12, Building N232, Rm. 146, P.O. Box 1,
Moffett Field, CA 94035-0001, USA}
\altaffiltext{6}{Max-Plank-Institut f\"ur Radioastronomie, Auf dem
H\"ugel 69, D-53121 Bonn, Germany}
\altaffiltext{7}{NRAO, 520 Edgemont Road, Charlottesville, VA, 22903,
USA}

\begin{abstract}

We report new results based on the analysis of the SMA and
CARMA observations of NGC 7538\,IRS\,1 at 1.3 and
3.4 mm with sub-arcsec resolutions. With angular resolutions
$\sim$ 0\farcs7, the SMA and CARMA observations show that the
continuum emission at 1.3 and 3.4 mm from the hyper-compact
\ion{H}{2} region IRS\,1 is dominated by a compact source
with a tail-like extended structure to the southwest of IRS\,1.
With a CARMA B-array image at 1.3 mm convolved to 0\farcs1,
we resolve the hyper-compact \ion{H}{2} region into two components:
an unresolved hyper-compact core, and a north-south extension
with linear sizes of $<270$ AU and $\sim$2000 AU, respectively.
The fine structure observed with CARMA is in good agreement with
the previous VLA results at centimeter wavelengths, suggesting that
the hyper-compact \ion{H}{2} region at the center of IRS\,1 is
associated with an ionized bipolar outflow. We image the
molecular lines OCS(19-18) and CH$_3$CN(12-11) as well as
$^{13}$CO(2-1) surrounding IRS\,1, showing a velocity gradient
along the southwest-northeast direction. The spectral line profiles
in $^{13}$CO(2-1), CO(2-1), and HCN(1-0) observed toward IRS\,1 show broad
redshifted absorption, providing evidence for gas infall
with rates in the range of $3-10\times10^{-3}$ M$_\odot$ yr$^{-1}$
inferred from our observations.

\end{abstract}

\section{Introduction}

NGC 7538 IRS 1 is a hyper-compact (HC) \ion{H}{2} region located at a
distance of 2.65 kpc \citep{mosc09}. Analysis of the radio and IR data
suggested that IRS 1 is an O6/7 star with a luminosity of a few times of
10$^5$ L$_\odot$ \citep{will76,akab05}. High-resolution
observations of IRS 1 with the Very Large Array (VLA) showed two lobes of
free-free emission within the central 2\arcsec\, region, located north
and south of the emission peak of the HC \ion{H}{2} region
\citep{camp84,gaum95,sand09}.
The centimeter emission has been interpreted as a north-south ionized
jet/outflow, while the large-scale outflow in CO appears to be along the
northwest-southeast direction \citep[e.g.][]{scov86,kame89,davi98,qiu11}.
At shorter wavelengths, a northeast-southwest elongation of dust emission
was observed in the mid-IR \citep{debu05} and at 0.87 mm, and
several molecular lines show similar velocity gradients
\citep{brog08,klaa09,beut12}. In addition, a northwest-southeast linear
structure of methanol masers with a velocity gradient of $\sim$0.02 km
s$^{-1}$ AU$^{-1}$ was found by \cite{mini98,mini00}. The kinematics
from the methanol maser emission was modeled as a circumstellar disk
\citep{pest04} with a major axis in the northwest-southeast direction
which is parallel
to the axis of the large-scale outflow. Alternatively, the linear
structure might be evident that the methanol masers trace the inner
wall of an outflow instead of a disk \citep{pest04}. To reconcile the
various models from the previous observations in CO, methanol masers,
radio continuum and mid-IR emission, \cite{krau06} proposed a precessing
outflow model to interpret the observations.

In order to investigate the astrophysical process in the inner region of
NGC\,7538 IRS 1, we carried out high-angular resolution observations with
the Submillimeter Array (SMA)\footnote{The Submillimeter Array is a joint
project between the Smithsonian Astrophysical Observatory and the Academia
Sinica Institute of Astronomy and Astrophysics and is funded by the
Smithsonian Institution and the Academia Sinica.} at 1.3 mm and
the Combined Array for Research in Millimeter-wave
Astronomy (CARMA) at 1.3 and 3.4 mm.
SMA archival data were also used in our analysis of the molecular lines
and continuum emission from IRS\,1.
We present the results of the 1.3-mm and 3.4-mm continuum emission in
Section 3 and molecular lines, including OCS(19-18), CH$_3$CN(12-11),
CO(2-1), $^{13}$CO(2-1) and HCN(1-0), in Section 4. The implications of
the results are discussed in Section 5. Section 6 summarizes the conclusions.

\section{Observations and Data Reductions}

\subsection{SMA Data}

SMA observations of NGC 7538 IRS 1 in the very-extended (VEX) configuration
were made on 2008 August 5 with LO frequency $\nu_{\rm LO}$ = 226.239 GHz,
giving band-center frequencies of 221 GHz for lower side band (LSB) and
231 GHz for upper side band (USB).
Eight antennas were used in the observations with the longest projected baseline
339 $k\lambda$. The spectral resolution was 0.81 MHz, which
corresponds to a velocity resolution of $\sim$1.1 km s$^{-1}$.
The total on-source integration time was 82 minutes.
The average system temperature was 180 K.

\begin{figure*}[]
  \centering
  \includegraphics[width=12.75cm, angle=-90, origin=c]{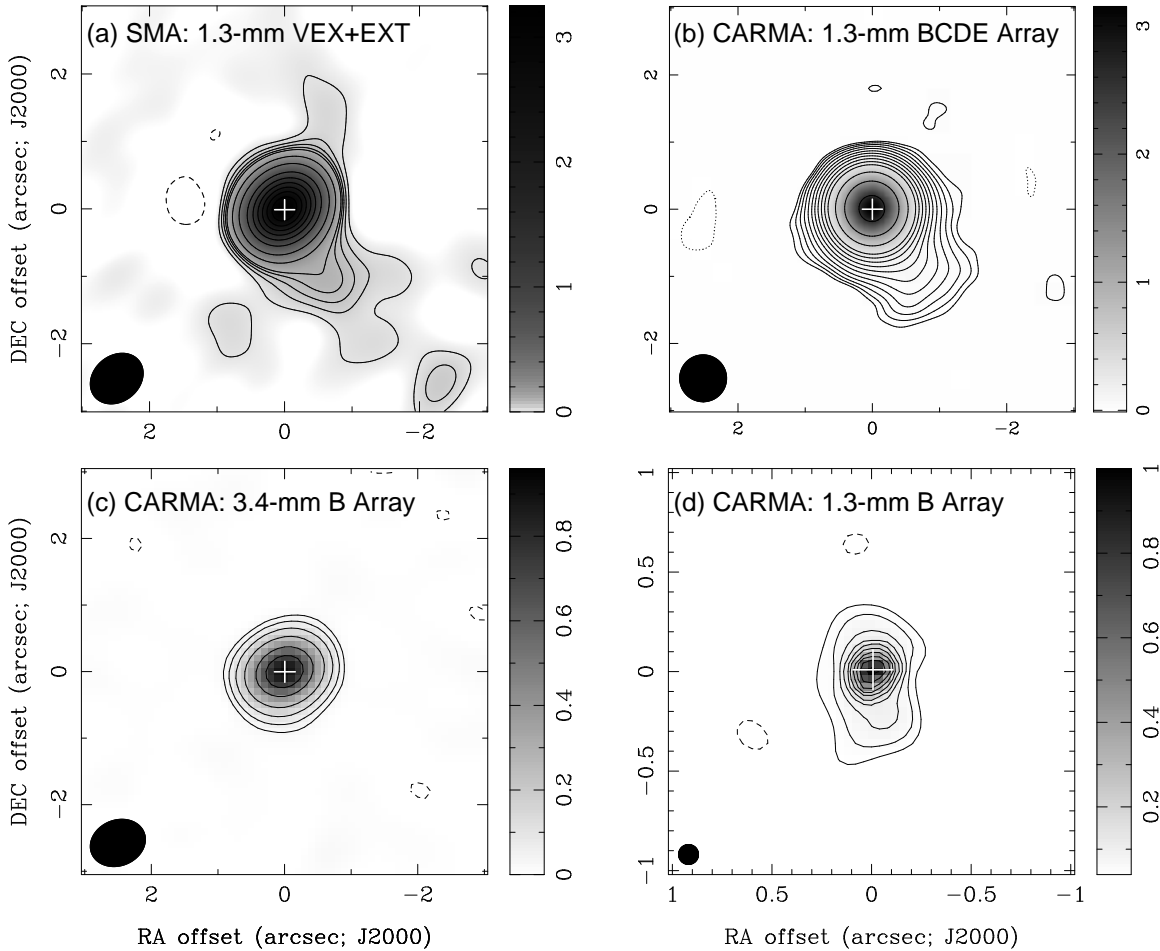}
  \vspace{-0.25cm}
\caption{(a) The SMA 1.3-mm continuum image made with the
line-free channels by combining LSB and USB data from both VEX and EXT
array observations. The contours are (--5, 5, 7, 10, 14, 20, 28, 40, 56,
80, 112, 160, 226, 320, 452, 640, 905, 1280)$\times\sigma$, and $\sigma=4$
mJy beam$^{-1}$. The FWHM beam size is 0\farcs8$\times$0\farcs6,
P.A.$=-45$\arcdeg.
(b) The CARMA 1.3-mm continuum image with lower-resolution made by
combining B, C, D and E array configuration data, convolved to a circular
beam (FWHM) of 0\farcs7, with rms noise, $\sigma=2.5$ mJy beam$^{-1}$.
The contours are (--4, 4, 5.7, 8, 11.3, 16, 23, 32, 45, 64, 91, 128, 256, 512
and 1028)$\times\sigma$.
(c) The CARMA 3.4-mm continuum image made from the CARMA B-array
observations with FWHM beam of 0\farcs8$\times$0\farcs7, P.A.$=-68$\arcdeg.
The contours are (--2, 4, 8, 16, 32, 64 and 96)$\times\sigma$. The rms
noise $\sigma=$ 7 mJy beam$^{-1}$.
In all the images the reference position is R.A. (J2000) =
23$^{\rm h}$13$^{\rm m}$45.$^{\rm s}$37, Dec. (J2000) =
61$\arcdeg$28$\arcmin$10\farcs43, which is the peak position of 43 GHz
continuum emission and marked by a plus. The FWHM of the synthesized
beam is indicated at the left-bottom corner of each image. The
greyscale of the wedges is in the units of Jy beam$^{-1}$.
(d) The CARMA 1.3-mm continuum image of the centeral region of IRS 1 with a
higher-resolution made with
the B-array data only, convolved to a circular beam (FWHM) of 0\farcs1,
and with rms noise, $\sigma=10$ mJy beam$^{-1}$. The contours are
(--4, 4, 5.7, 8, 11.3, 16, 23, 32, 45, 64, 91)$\times\sigma$.
}
\end{figure*}

\begin{figure*}
  \centering
  \includegraphics[width=5.6cm, angle=-90, origin=c]{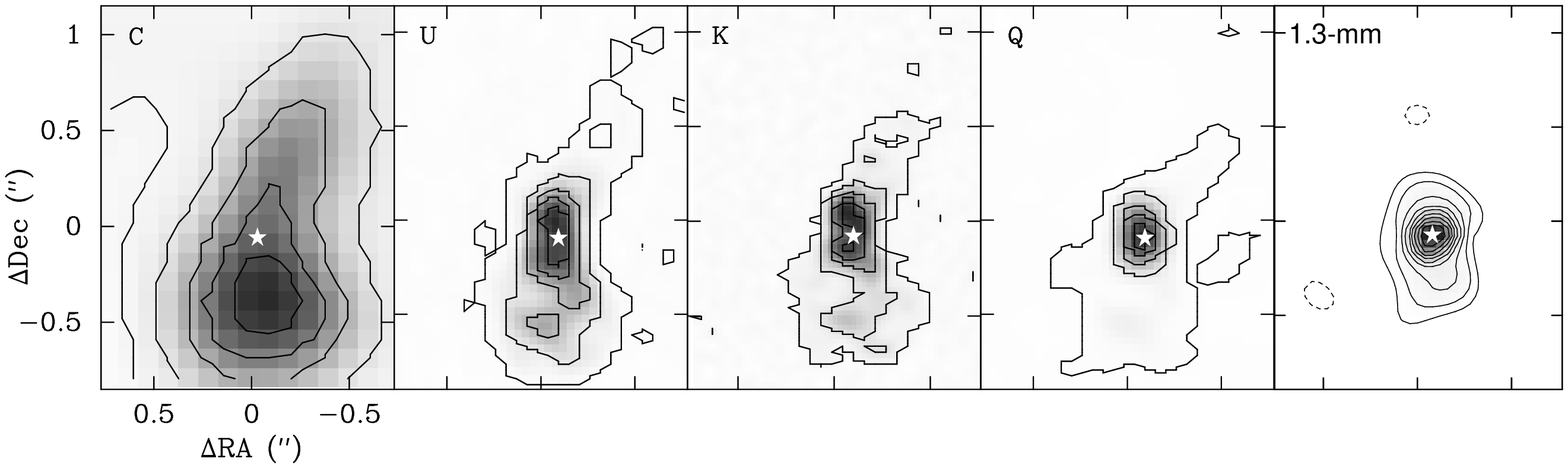}
  \vspace{-6.25cm}
\caption{The CARMA high-resolution image of IRS\,1 at 1.3 mm
is compared with the images observed with the VLA at 6, 2.0,
1.3 and 0.7 cm \citep{sand09}. Different from other figures
in this paper, this reference center is the same as the original
figure in \cite{sand09}. The white stars mark the position R.A. (J2000) =
23$^{\rm h}$13$^{\rm m}$45.$^{\rm s}$37, Dec. (J2000) =
61$\arcdeg$28$\arcmin$10\farcs43.}
\end{figure*}

NGC 7538 IRS 1 was also observed with the SMA in the extended (EXT)
configuration on 2005 September 11, with an LO frequency $\nu_{\rm LO}$
= 226.348 GHz. Six of the eight SMA antennas were used, giving the
longest projected baseline 133 $k\lambda$. A standard correlator
configuration was used with a spectral resolution of 0.33 MHz,
corresponding to a velocity resolution of 0.42 km s$^{-1}$.
The total on-source integration time was 309 minutes, and the average
system temperature was 105 K.

The data reduction and analysis were made in MIRIAD \citep{stw95}
following the reduction instructions for SMA data\footnote{\it
http://www.cfa.harvard.edu/sma/miriad}. For the EXT array data set (2005),
system temperature corrections were applied to remove the atmospheric attenuation.
Antenna-based bandpass solutions were calculated over three time intervals
(UT 4:45-5:40 for Neptune, UT
6:00-6:30 and 14:00-14:45 for 3C454.3) to interpolate the corrections
for the bandpass ripples due to instrument defects.
After that, baseline-based and spectral-window based residual errors in bandpass
were further corrected using the point source 3C454.3. The time-dependent
gains were derived and interpolated from two nearby QSOs J0102+584 and J2202+422;
the complex gain corrections were applied to the target source.
Finally, the flux-density scale was bootstrapped from
Ceres using the SMA planetary model assuming a flux-density of 0.95 Jy
and an angular size of 0\farcs43.
The reduction of the VEX array data is similar, and the calibrators used in
the data reduction for the EXT and VEX configuration
data are summarized in Table 1.

In order to image the continuum emission and the molecular lines
in the IRS\,1 region
with high angular resolution and good uv coverage, we combined the EXT
(2005) and VEX (2008) datasets. The EXT and VEX visibility datasets have
an offset of 0\farcs7 in pointing position. Using {\it UVEDIT} in
MIRIAD, we shifted the two data sets to a common phase reference center
at the continuum peak position determined from the EXT (2005) data set.
Then, the residual phase errors in the continuum were further corrected
using the self-calibration technique with a point-source model at the
position of the radio compact source in IRS\,1 determined from the VLA
image at 43 GHz \citep{sand09}. The self-calibration solutions determined
from continuum data were also applied to the line data sets.

The continuum data sets were created by averaging all the line-free
channel data using {\it UVLIN}. Combining the USB and LSB datsets of
the EXT and VEX array configurations, the continuum image at
1.3 mm was constructed.
The continuum-free line visibility data sets were created by subtracting
the continuum level determined from a linear interpolation from the
line-free channels using {\it UVLIN} in MIRIAD.

In addition to the 1.3-mm results, we also included the SMA data of
NGC 7538 IRS 1 at 345 GHz which were observed in 2005 \citep{brog08}.
The FWHM beam sizes for the final line images are about 2\arcsec. We
analyzed the relatively optically thin lines of species CH$_3$OH and
$^{13}$CH$_3$OH to determine the systemic velocity of the IRS 1 system.

\subsection{CARMA Data}

\subsubsection{1.3 mm}

High-resolution observations of IRS 1 were carried out on January 4
and 5, 2010 with CARMA in the B array configuration at 1.3 mm. The CARMA
observations used two 500 MHz bands in upper and lower sidebands of
LO1 with a total continuum bandwidth of 2 GHz. The data were reduced and
imaged in a standard way using MIRIAD software. The quasar J0102$+$584
was used as a gain and phase calibrator, and Uranus and MWC\,349 for flux
and bandpass calibration. The uncertainty in the absolute flux-density scale
was $\sim$20\%. The strong compact emission from IRS\,1 was used to
self-calibrate IRS\,1 with respect to the 43 GHz VLA image with a Gaussian
fit position R.A. (J2000) = 23$^{\rm h}$13$^{\rm m}$45.$^{\rm s}$37,
Dec. (J2000) = 61\arcdeg28\arcmin10\farcs4. The continuum image was
constructed by combining the two-500 MHz sidebands in a multi-frequency
synthesis (MFS) mosiac. The CARMA data with three different primary beams 
resulting from the 6.1 and 10.4 m antenna pairs were used in the mosaic imaging. 
The weighted mean observing frequency from multi-frequency synthesis is 222.2 GHz.

To verify the structure observed with the SMA, we used
the lower-resolution data from the CARMA observations in C, D and E
array configurations that were carried out between 2007 to 2010. The
setup of the corresponding observations are summarized in Table 1,
giving observing date, pointing positions, band information, array
configurations and calibrators of these observations. All of these
observations were made with two or three 500-MHz wide bands in upper and
lower sidebands with total bandwidths of 2 or 3 GHz. These continuum data
were calibrated following the standard procedure. The calibrated datasets
were split into single source datasets for the target source IRS 1.
With {\it UVEDIT}, the phase-centers of each dataset were shifted to the
common position of the IRS\,1 continuum source determined from the VLA
image at 43 GHz. Then, the residual complex gain errors in all the B,
C, D and E array data sets were corrected using self-calibration with
an initial model of a point source at the phase center.
The structure seen in the B-array image which was self-calibrated with the model
from the 43-GHz VLA image (as discussed in the above paragraph) was confirmed
from this self-calibration procedure using a point-source as an initial model.

\begin{figure*}[]
  \centering
  \includegraphics[width=16cm, angle=0, origin=c]{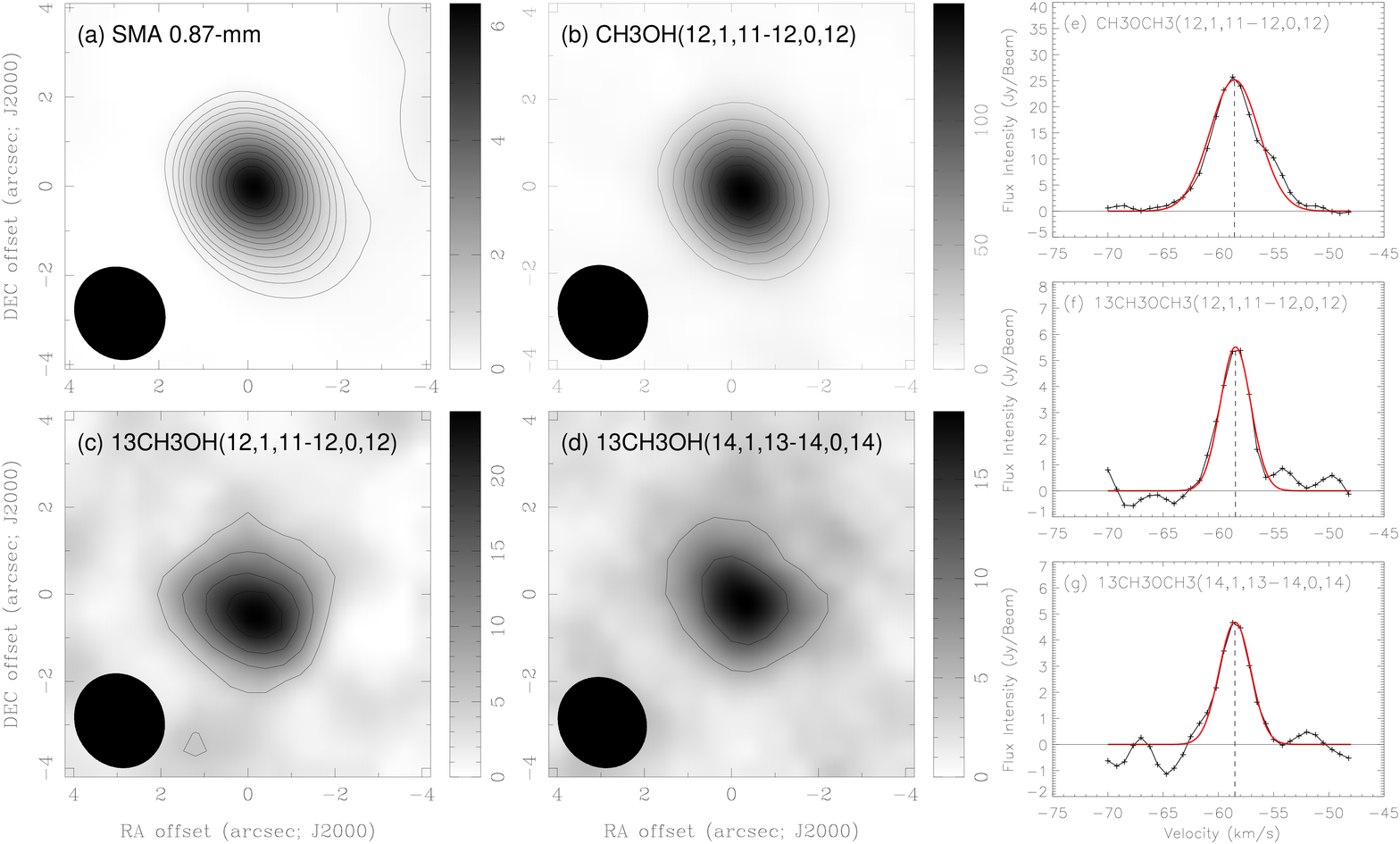}
  \caption{(a) The SMA 0.87-mm continuum image with a FWHM beam size of
  2\farcs1$\times$2\farcs0, P.A. $=38$\arcdeg. The contours are (--1, 1,
  2, 3, 4, 6, 8, 10, 12, 15, 18, 21, 24, 28)$\times$5$\sigma$, and the
  rms noise is $\sigma=45$ mJy beam$^{-1}$.
  (b) The integrated intensity map of the line CH$_3$OH(12,1,11-12,0,12)
  with a FWHM beam size of 2\farcs2$\times$2\farcs0, P.A. $=29$\arcdeg.
  The contours are (-1, 1, 2, 3, 4, 5, 6, 7, 8)$\times$10$\sigma$, and the
  rms noise is $\sigma=1.7$ Jy beam$^{-1}$ km s$^{-1}$.
  (c) The integrated intensity map of the line $^{13}$CH$_3$OH(12,1,11-12,0,12)
  with a FWHM beam size of 2\farcs2$\times$2\farcs0, P.A. $=31$\arcdeg.
  The contours are (-1, 1, 2, 3, 4, 5, 6, 7, 8)$\times$3$\sigma$, and the rms
  noise is $\sigma=1.7$ Jy beam$^{-1}$ km s$^{-1}$.
  (d) The integrated intensity map of the line $^{13}$CH$_3$OH(14,1,13-14,0,14)
  with a FWHM beam size of 2\farcs1$\times$2\farcs0, P.A. $=38$\arcdeg.
  The contours are (-1, 1, 2, 3, 4, 5, 6, 7, 8)$\times$3$\sigma$, and the rms
  noise is $\sigma=2.0$ Jy beam$^{-1}$ km s$^{-1}$.
  (e) - (g) The spectra of CH$_3$OH(12,1,11-12,0,12),
  $^{13}$CH$_3$OH(12,1,11-12,0,12) and $^{13}$CH$_3$OH(14,1,13-14,0,14) at
  the peak position of the continuum emission. The red lines in (e) - (g)
  are the results of Gaussian fitting for the line profiles.}
\end{figure*}

\begin{figure*}[]
  \centering
  \includegraphics[width=6.5cm, angle=90, origin=c]{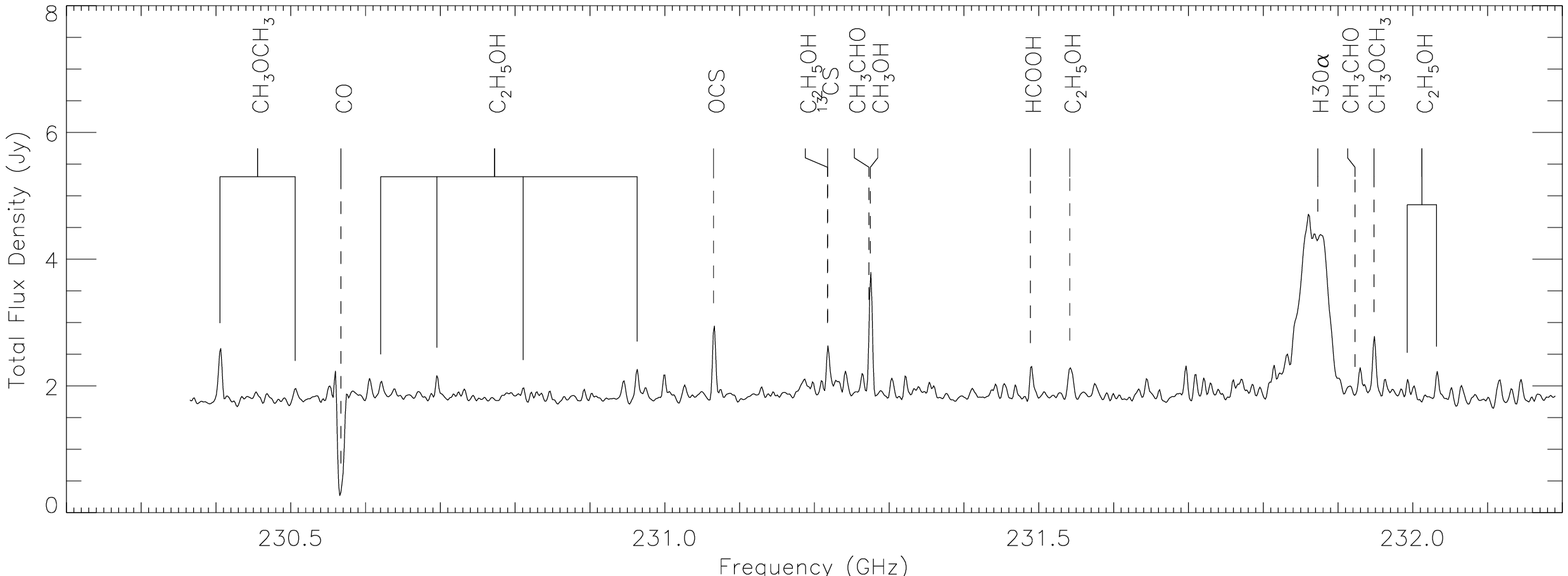}
  \\
  \vspace{-5.5cm}
  \includegraphics[width=6.5cm, angle=90, origin=c]{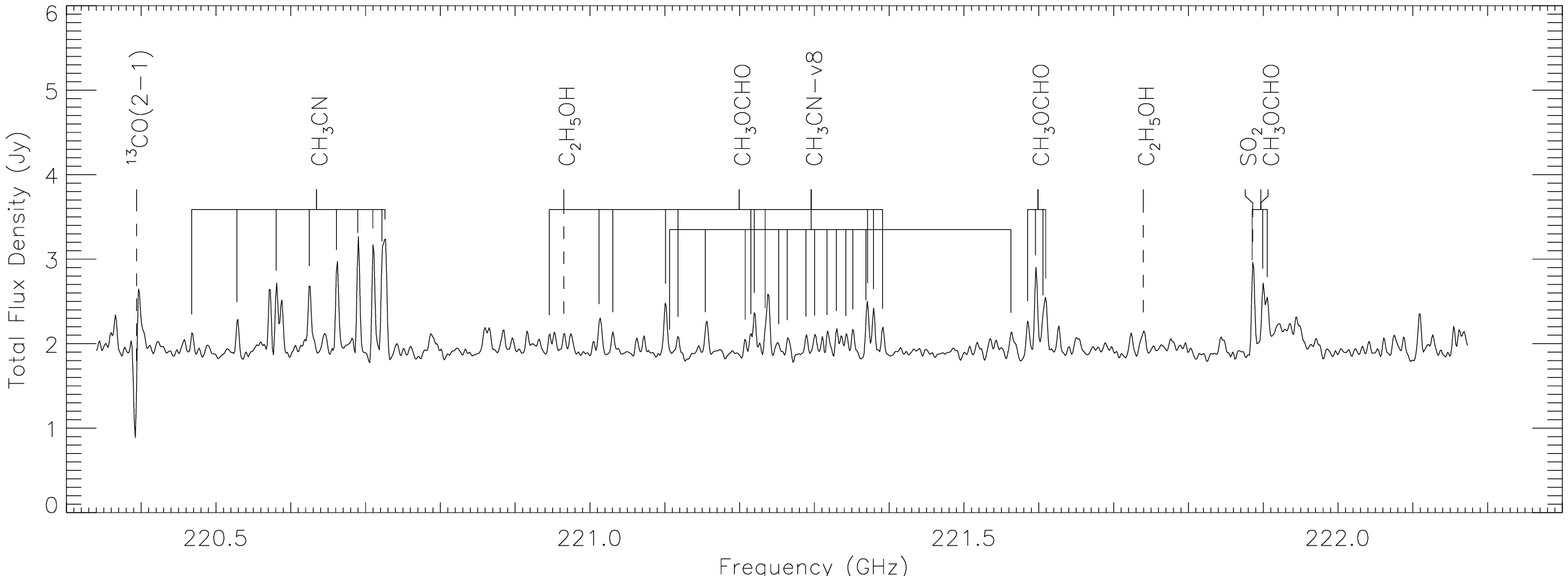}
  \vspace{-5.25cm}
\caption{The spectra toward IRS\,1 that made from the upper (top)
  and lower (bottom) side bands based on the observations with the SMA
  on September 11, 2005 in the extended array configuration. The spectra
  have been converted to the rest frame at $V_{\rm sys}=-59.0$ km s$^{-1}$. The
  identifications for the major molecular and hydrogen recombination
  lines are labeled.}
\end{figure*}

\subsubsection{3.4 mm}

We also carried out observations at 89\,GHz with the CARMA telescope
in the B array configuration with 15-antennas on December 28, 2010. We
used a correlator setup with 16 spectral windows (including upper and
lower sidebands), four with 500\,MHz bandwidth (39 channels in total and
12.5\,MHz for each channel), four with 250\,MHz (79 channels in total
and 3.225\,MHz for each channel) and the remaining eight windows with
64\,MHz bandwidth (255 channels in total and 0.244\,MHz for each channel).

The data were calibrated and imaged following the standard procedure
for CARMA data in MIRIAD.
The HCN(1-0) line at $\nu_0=$88.6318\,GHz was included in one of
the narrow-band windows. The continuum data set was made from
the line-free channels in this window. The residual phase errors
were corrected using the self-calibration technique.
The position of the IRS\,1 continuum source at 3.4 mm
was offset (--0\farcs18,--0\farcs10) from the SMA position at 1.3 mm,
which is within the positional uncertainty in the SMA image.
We binned two channels together to make a line image of HCN(1-0)
with the same FWHM beam size as the continuum image.

A log summarizing the observations, uv datasets and images is
given in Table 1.

\section{Structure of Continuum Emission from IRS\,1}

With an angular resolution of $\sim$0\farcs7, both the SMA (Figure 1a)
and CARMA (Figure 1b) show extended emission with a peak flux density of
0.12$\pm$0.02 Jy beam$^{-1}$, located $\sim$1\arcsec SW of the
unresolved component in IRS\,1. Lacking centimeter counterparts,
the extended continuum emission is likely to correspond to thermal
dust emission, which could contain additional dust cores in the IRS\,1
region. It is also plausible that the SW extension corresponds to a
dust component which is located along the interface between the
ionized and molecular medium and is compressed by the ionized outflow.

The CARMA image at 3.4 mm (Figure 1c) shows a point-like
source with no significant detection of the SW extension. A total
flux density of 0.93$\pm$0.10 Jy was determined for the central unresolved core.
Assuming a flat power-law ($\beta\sim0.5$ for dust emissivity), a peak
flux density of 0.01 Jy beam$^{-1}$ is extrapolated from the
1.3-mm peak flux density of the SW extension, which is below the
3$\sigma$ limit of the 3.4 mm-image.

The CARMA 1.3-mm image (Figure 1d), convolved to a circular beam of
0\farcs1, shows a central emission peak with a flux density of
1.0$\pm$0.1 Jy beam$^{-1}$.
The innermost part of the HC~\ion{H}{2} region at 1.3 mm is un-resolved,
which could be the optically thick part of the ionized outflow
containing an accretion disk in IRS 1.
Because the optically thick free-free
emission dominates the 1.3-mm continuum, as well as the
limitation of the angular resolution, the postulated ionized disk in
IRS\,1 has not been detected.
In the CARMA 1.3-mm image (Figure 1d), extensions along the south and
north directions from the bright point source has been revealed with
a scale of 0\farcs8 above a noise level of 4 $\sigma$. The emitting
lobes appear to be from the inner portion of the ionized outflow.

Previous studies show that the continuum emission from IRS 1\,is
dominated by the free-free emission from a bipolar ionized outflow
at frequencies lower than 300 GHz \citep{sand09}.
Figure 2 compares the CARMA 1.3-mm image with previous
VLA images. At wavelengths
$\lambda\ge1.3$\,cm, a dark lane divides two emission peaks in the
center of the IRS\,1 region, which is probably due to the
self-absorption in the ionized gas. At shorter wavelengths,
$\lambda\le 7$\,mm, a single emission peak appears, located at
R.A. (J2000) = 23$^{\rm h}$13$^{\rm m}$45.$^{\rm s}$37,
Dec. (J2000) = 61\arcdeg28\arcmin10\farcs43. Hereafter, we take
this position as the reference to register both the SMA and CARMA
images.

The properties of the continuum sources in the IRS\,1 region are
summarized in Table 2.


\begin{figure*}[]
  \centering
  \includegraphics[width=10cm, angle=-90]{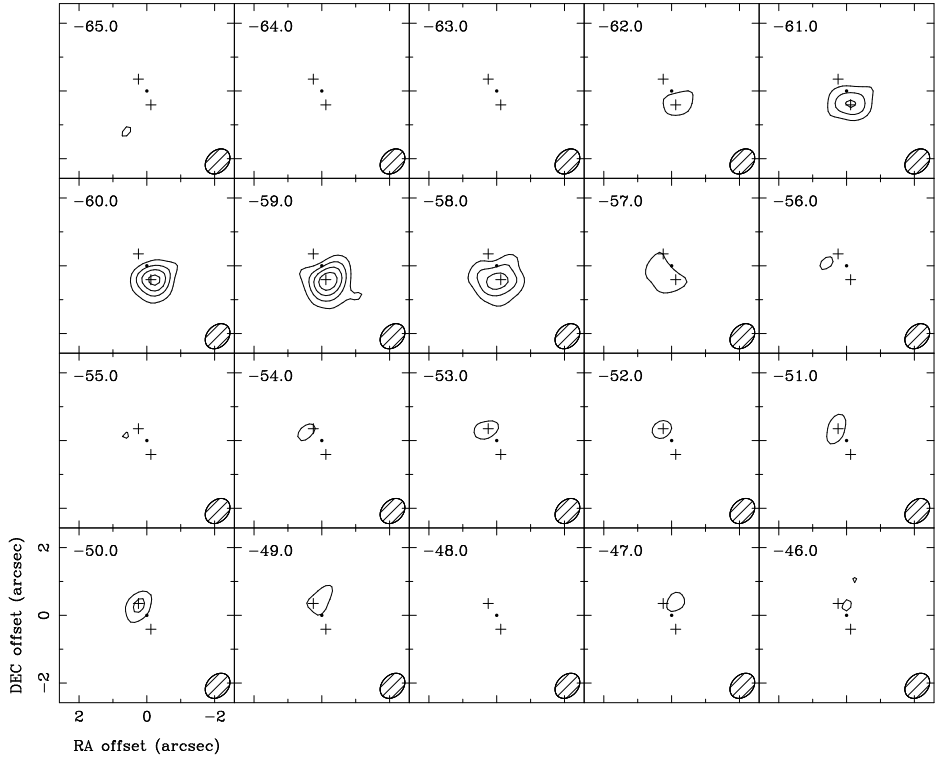}
  \caption{The channel maps of the OCS(19-18) lines.
  The contours are (--1, 1, 2, 3, 4, ...)$\times$5$\sigma$, and
  1 $\sigma=0.06$ Jy beam$^{-1}$. The FWHM beam size is
  0\farcs8$\times$0\farcs6, P.A.$=-43\arcdeg$. IRS\,1 is located
  at the phase center (marked by filled black circles), and
  the intensity peaks of the SW and NE components are marked by
  plus signs.}
\end{figure*}

\section{Molecular lines}

\subsection{Systemic Velocity}

At 1.3-mm wavelength, the SMA data for NGC\,7538\,IRS\,1 show numerous
strong molecular emission lines along with a few absorption features.
A large portion of the emission is from the molecular lines with high
excitation energy levels (e.g. OCS, CH$_3$CN, CH$_3$OCH$_3$ and C$_2$H$_5$OH),
showing a strong emission peak at a radial velocity of $-59.5$ km s$^{-1}$
with a scatter of about one channel width ($\sim$0.5 km s$^{-1}$). This
value is close to the velocity ($-59.7\pm0.3$ km~s$^{-1}$) observed
in the mid-infrared by \citet{knez09}. In addition, we further investigated the
0.86-mm SMA data and analyzed the CH$_3$OH(12,1,11-12,0,12),
$^{13}$CH$_3$OH(12,1,11-12,0,12) and $^{13}$CH$_3$OH(14,1,13-14,0,14) lines.
The integrated intensity images and line profiles are shown in Figure 3,
and the Gaussian fitting to the line profiles are summarized in Table 3. The
Gaussian fitting shows an average peak velocity of $-58.6\pm0.2$ km
s$^{-1}$ for these lines. These lines are with higher excitation energy than
the ones observed in the 1.3-mm SMA data, as well as lower optical depths
(for the two rare isotopic ones). On the other hand, since the $\sim$2\arcsec
angular resolution of the 0.86-mm observations is poor compared to the 1.3-mm
observations, it is possible that the 0.86-mm emission includes
the kinematics of surrounding medium (e.g., outflow, infall). Therefore,
considering the results of all the observations we adopt
$V_{\rm sys}=-59.0\pm0.5$ km s$^{-1}$, as the systemic velocity of the IRS\,1
system in the following analysis.

\begin{figure*}[]
  \begin{minipage}[]{0.6\linewidth}
  \includegraphics[width=9cm, angle=-90, origin=c]{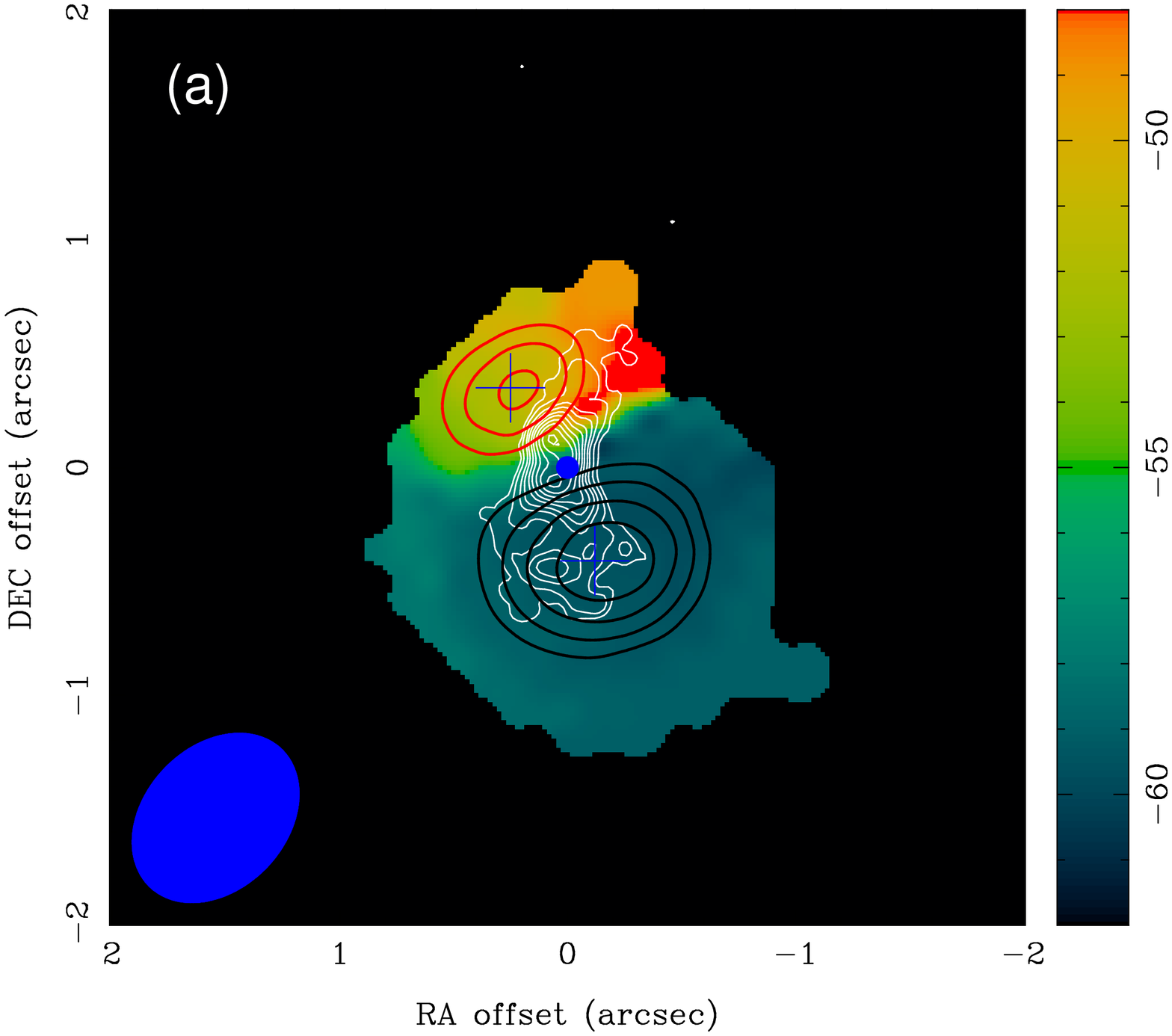}
  \end{minipage}
  \begin{minipage}[]{0.3\linewidth}
  \vspace{-1cm}
  \includegraphics[width=6.25cm, angle=0, origin=c]{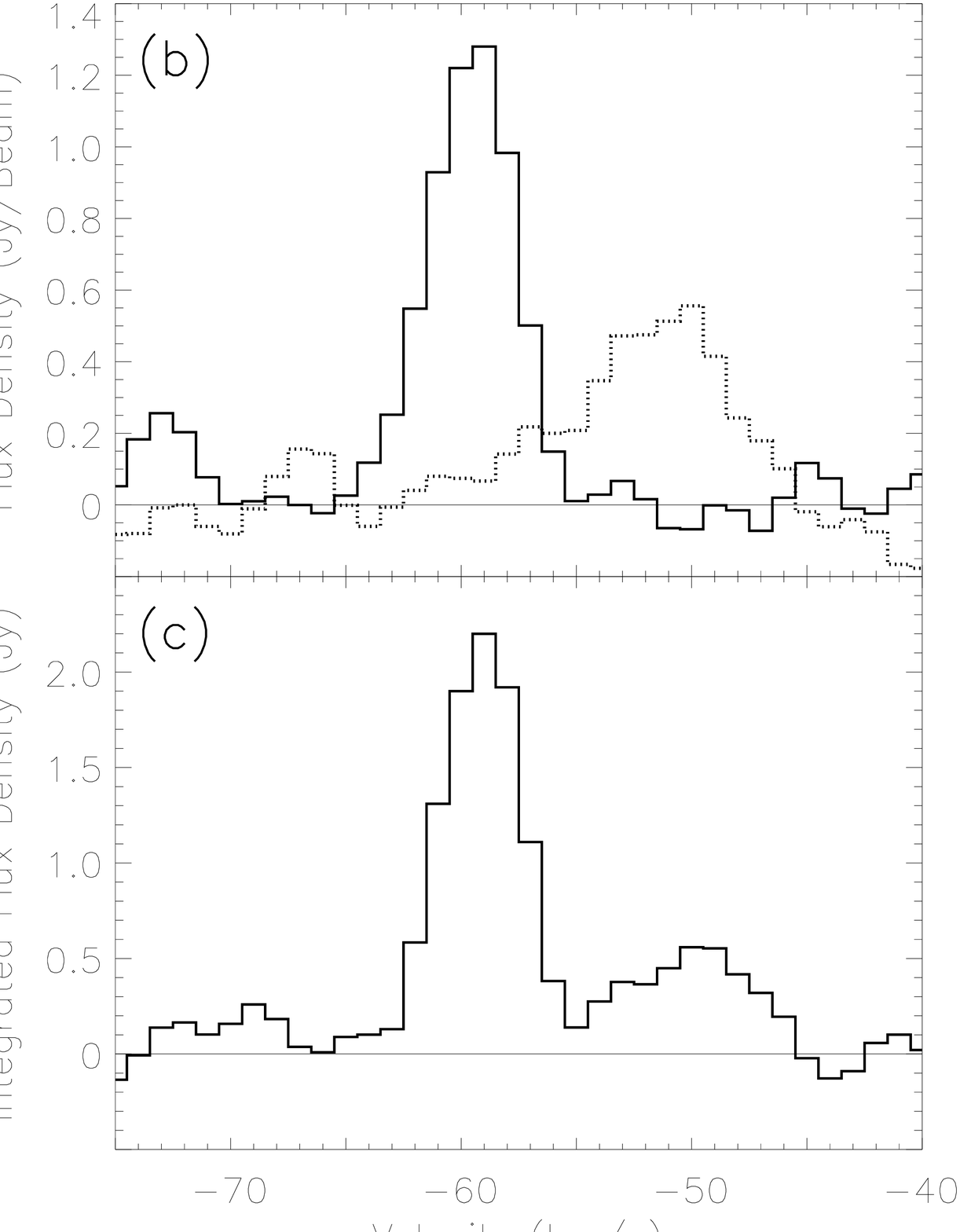}
  \end{minipage}
 \vspace{-0.5cm}
  \caption{(a) The integrated intensity maps of the OCS(19-18)
  components IRS\,1 NE (red) and SW (black) superimposed on
  the centroid velocity image (color background). The velocity
  ranges of the integrations are ($-54.0$, $-47.0$) km s$^{-1}$
  and ($-62.5$, $-57.0$) km s$^{-1}$ for IRS 1 NE and SW, respectively.
  The contour levels are (2.3, 2.8, 3.3) and (3.0, 3.8, 4.6, 5.4)
  Jy beam$^{-1}$ km s$^{-1}$ for IRS\,1 NE and SW, respectively.
  The rms noise $\sigma$ is 0.15 Jy beam$^{-1}$ km s$^{-1}$.
  The FWHM beam size is 0\farcs8$\times$0\farcs6, P.A. $=-45$\arcdeg.
  The unit of the wedge scaling is km s$^{-1}$.
  The white contours present the 1.3-cm continuum emission from
  the VLA observations \citep[re-produced according to ][]{gaum95}.
  IRS\,1 is located at the phase center, marked by blue
  dot. Plus signs mark the peak positions of IRS 1 NE and SW in
  OCS(19-18). (b) The line profiles of OCS(19-18) at the peak positions of IRS
  1 SW (solid line) and 1 NE (dashed line). (c) The line profile of OCS(19-18)
  integrated over the entire IRS 1 region.
  }
\end{figure*}

\subsection{Line Identifications of the 1.3-mm SMA Data}

Figure 4 shows the spectra from the 1.3-mm SMA data observed
in the EXT configuration, which are corrected for the
systemic velocity, $V_{\rm sys}=-59.0$ km s$^{-1}$.
Numerous molecular lines were identified on the basis of the molecular
line surveys toward Sgr B2 \citep{numm98} and Orion \citep{sutt85}, as
well as other available line catalogs. The names of the identified
molecules are labeled in Figure 4.

With no significant blending with other lines, OCS(19-18) and
multiple $K$ transitions of CH$_3$CN(12-11) included in both EXT and VEX
configuration observations were selected in our analysis to determine
the excitation condition and kinematics in the IRS\,1 region.


\subsection{OCS(19-18)}

The channel maps of the OCS(19-18) line show two significant velocity
components in OCS(19-18) emission (see Figure 5). The main OCS(19-18) emission
component in the velocity range from $-62.5$ to $-57.0$ km s$^{-1}$ is
referred to as IRS\,1 SW as it is located to the south and southwest of
the HC \ion{H}{2}
region IRS\,1, while the fainter, less extended redshifted
component, hereafter IRS\,1 NE, is located to the northeast of IRS\,1
ranging from $-52.0$ to $-47.0$ km~s$^{-1}$. No significant OCS(19-18)
emission is present in the velocity range between $-56.5$ and $-53.0$
km~s$^{-1}$. This velocity gap in emission is redshifted with respect
to $V_{\rm sys}=-59.0$ km s$^{-1}$.

Figure 6a shows the integrated line intensity of the two
components SW and NE. We fitted the two components with
2-D Gaussians as well as their line profiles with multiple Gaussians.
The fitting results are summarized in Table 4. The
peak position of IRS\,1 SW is offset
($-0\farcs13\pm0\farcs03$, $-0\farcs41\pm0\farcs03$) with
respect to IRS\,1 while the offset of IRS\,1 NE is
($0\farcs25\pm0\farcs03$, $0\farcs35\pm0\farcs02$). The total
integrated line flux of OCS(19-18) for IRS\,1 SW ($10.1\pm0.7$ Jy
km s$^{-1}$) is about three times of the value for
IRS\,1 NE ($3.8\pm0.6$ Jy km s$^{-1}$).

Figure 6b shows the line profiles of the OCS(19-18) spectra at the
peak positions of IRS\,1 SW and NE,
and Figure 6c shows the spectral profile integrated over the entire
IRS\,1 region. Gaussian fits to the observed spectral profiles give a
radial velocity $V_{\rm LSR}=-59.5\pm0.1$ km s$^{-1}$ with a line width
$\Delta V_{\rm FWHM}=4.4\pm0.1$ km s$^{-1}$ for IRS\,1 SW, and
$V_{\rm LSR}=-51.5\pm0.3$ km s$^{-1}$ and
$\Delta V_{\rm FWHM}=7.2\pm0.6$ km~s$^{-1}$ for the weaker
emission feature (IRS\,1 NE) (see Table 4).
The centroid velocity of IRS 1 SW is close to
$V_{\rm sys}=-59.0$ km s$^{-1}$ used in this paper,
suggesting that ambient gas might dominate this component.
The spectrum of IRS 1 NE is not a well-defined Gaussian shape,
and the large line width suggests that it is subject to
more dynamic interaction than IRS\,1 SW.


\begin{figure*}[]
  \centering
  \includegraphics[width=5.4cm, angle=0]{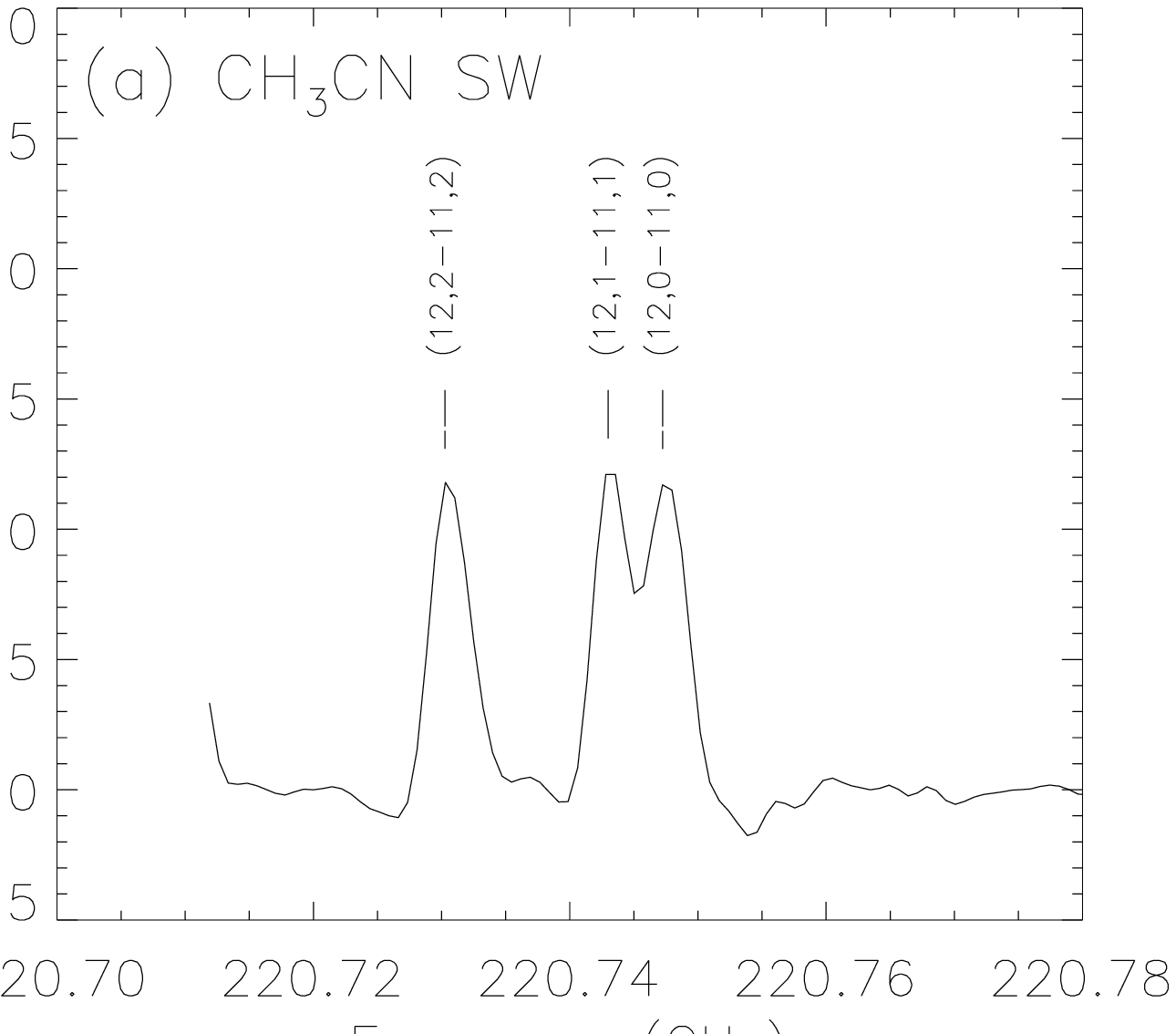}
  \hspace{0.8cm}
  \includegraphics[width=8.1cm, angle=0]{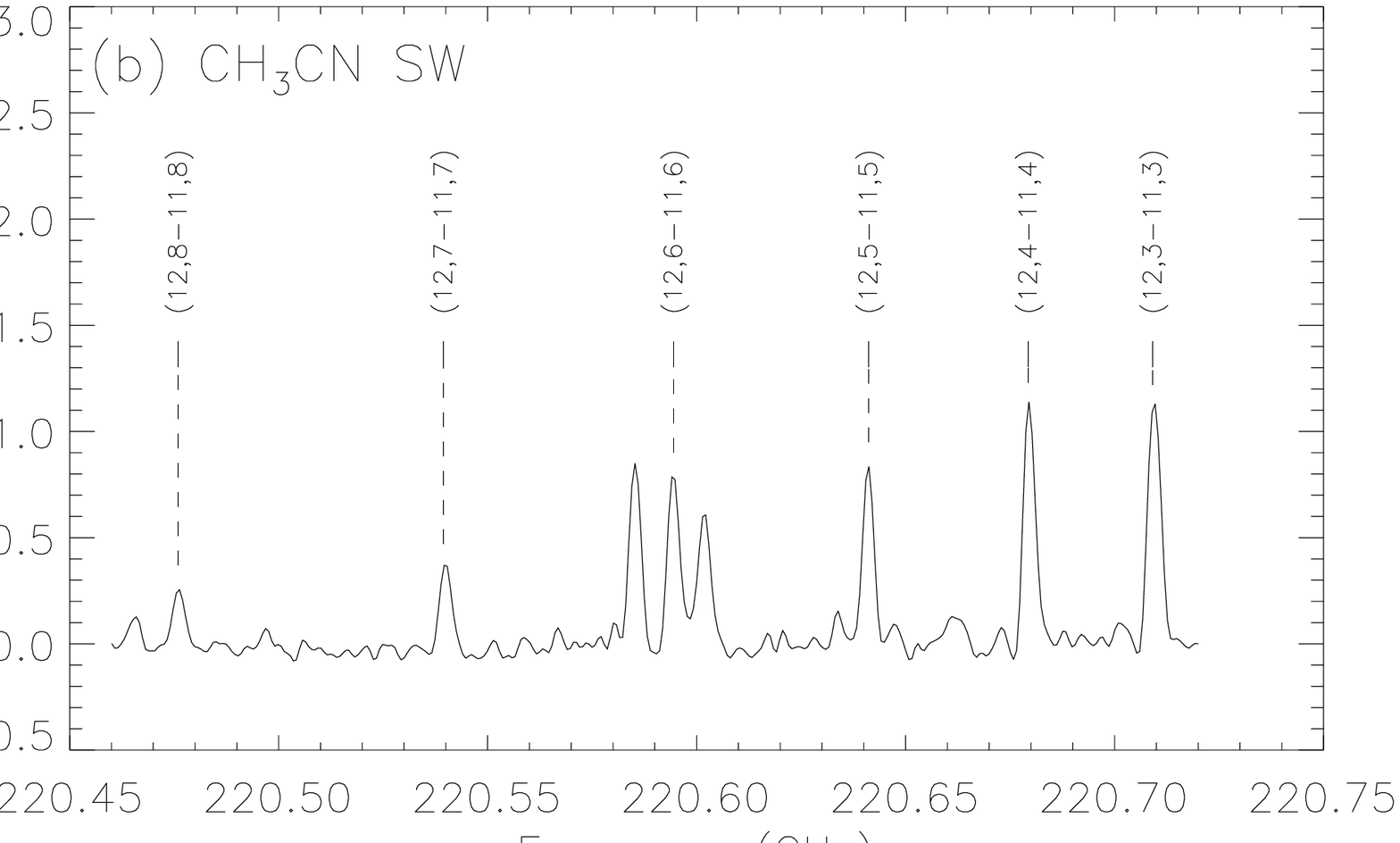}
  \\
  \includegraphics[width=5.4cm, angle=0]{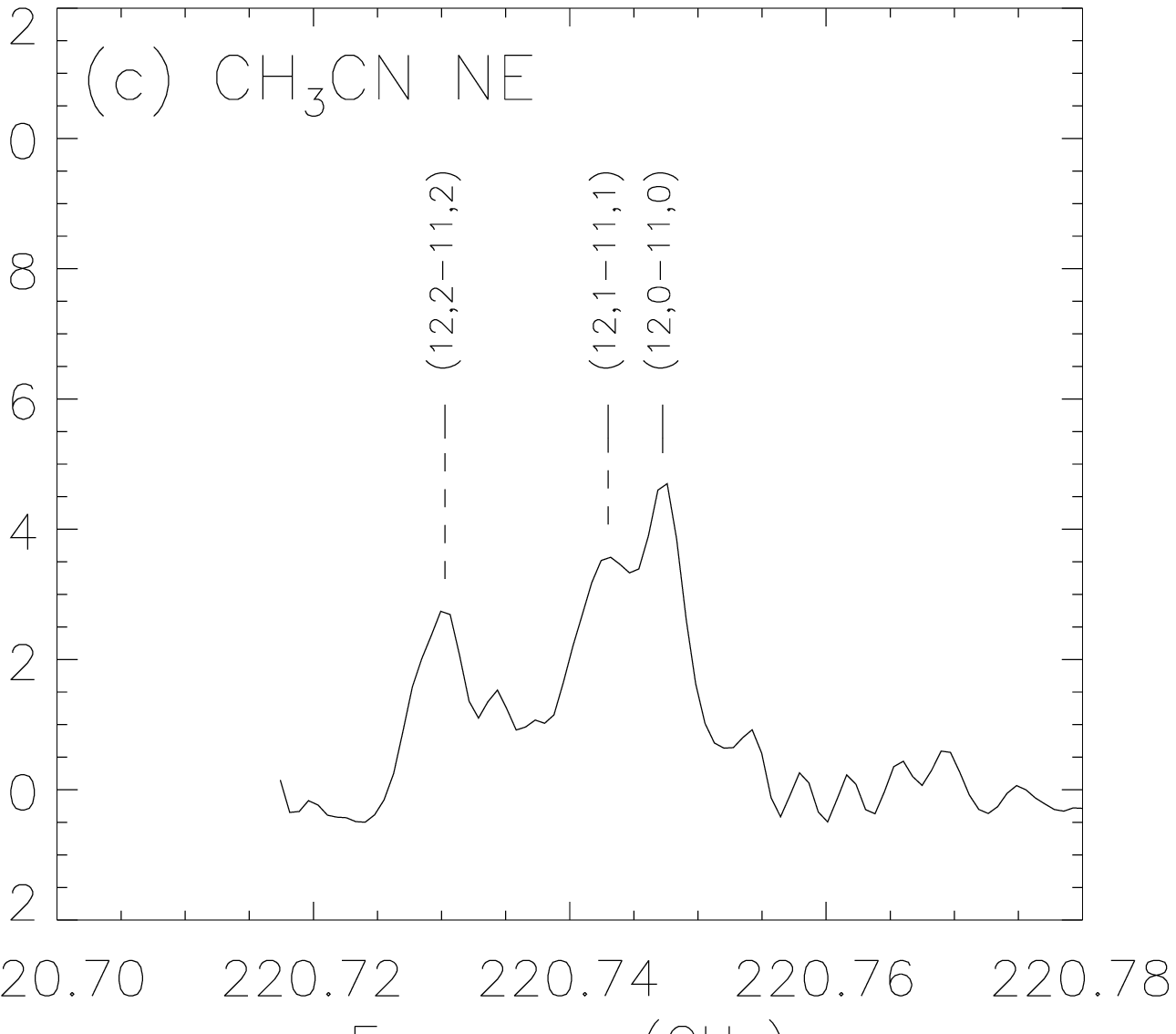}
  \hspace{0.8cm}
  \includegraphics[width=8.1cm, angle=0]{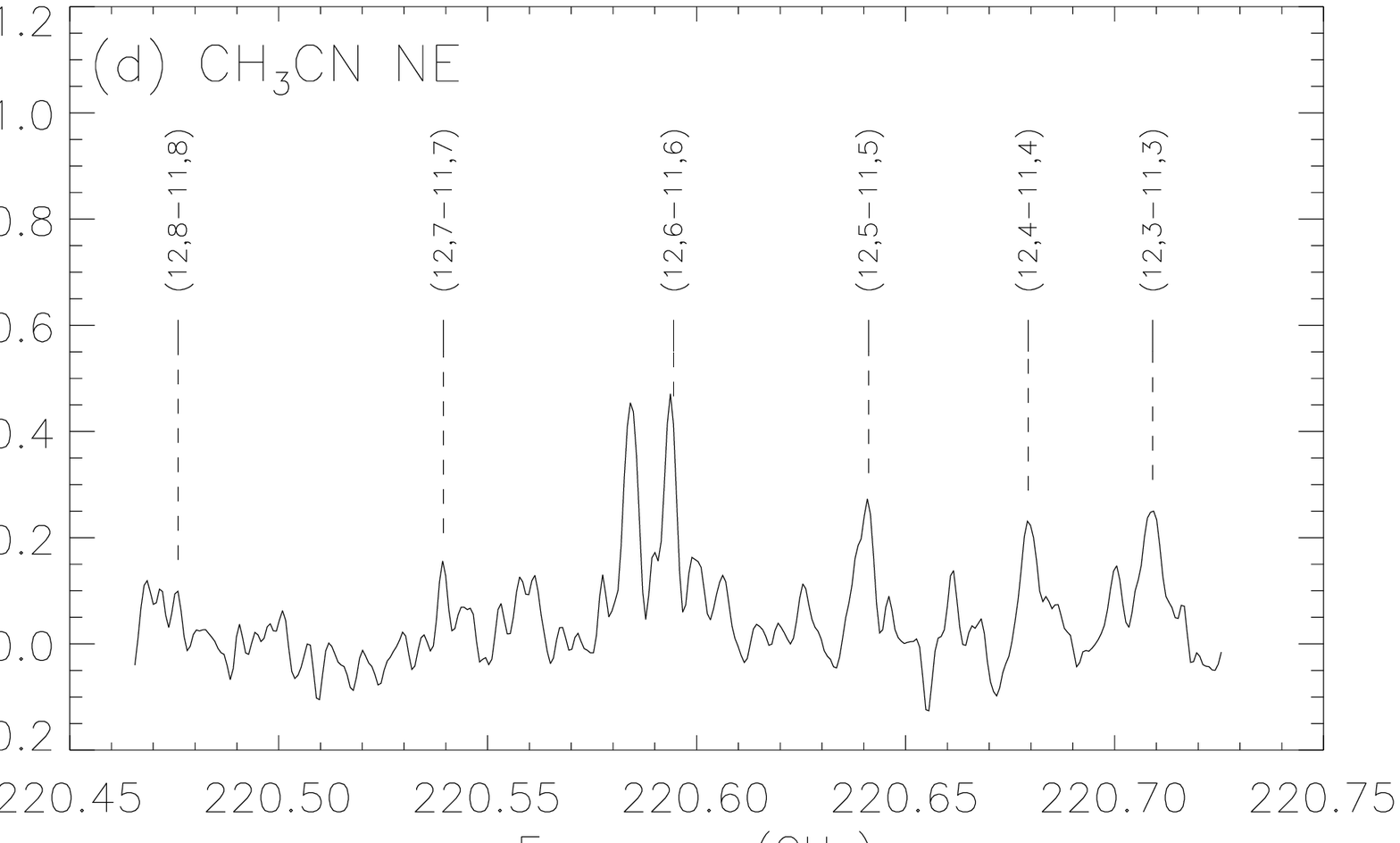}
  \vspace{0.5cm}
  \caption{(a)-(b) The line profiles of CH$_3$CN(12-11),
  $K=0,1,2$ and $K=3,4,5,6,7,8$ at the intensity peak position of
  IRS 1 SW. The spectra are converted to
  $V_{\rm LSR}=-59.5$ km s$^{-1}$. The blending due to
  component NE is neglectable at this position.
  (c)-(d) The line profiles of CH$_3$CN(12-11),
  $K=0,1,2$ and $K=3,4,5,6,7,8$ at the intensity peak position of
  IRS 1 NE. The spectra are converted to
  $V_{\rm LSR}=-51.5$ km s$^{-1}$. In (c) the transition $K=0$
  is blended with the transition $K=1$ from IRS 1 SW.
  }
\end{figure*}

\begin{figure*}[]
  \centering
  \includegraphics[width=9cm, angle=-90]{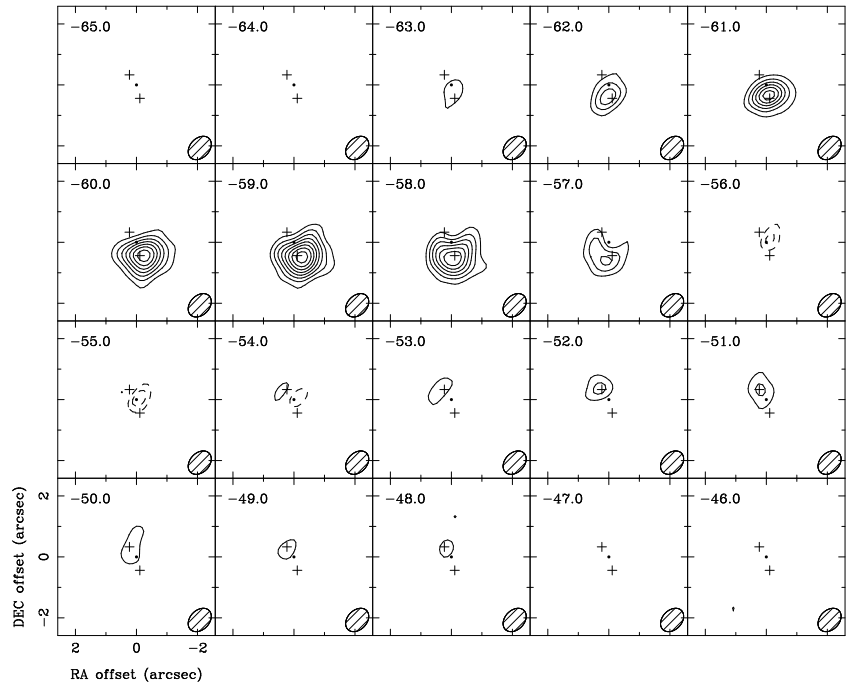}
  \caption{The channel maps of CH$_3$CN(12-11) were constructed
  by combining the visibility data of $K$=2-5 transitions
  in the LSR velocity domain, which is equivalent to
  averaging the line intensities from the $K$=2-5 transitions.
  The contours are (--1, 1, 2, 3, 4, ...)$\times$5$\sigma$,
  and 1 $\sigma=0.03$ Jy beam$^{-1}$. The FWHM beam size is
  0\farcs8$\times$0\farcs6, $\rm P.A.=-45$\arcdeg. IRS\,1 is located
  at the phase center (indicated by filled black circles), and
  the intensity peaks of the SW and NE components are marked by
  plus signs.}
\end{figure*}

\subsection{CH$_3$CN(12-11) Lines}

The CH$_3$CN(12-11) $K$-ladder with $K$=0-8, which covers
the frequencies ranging from $\nu_{K=8}$ (220.476 GHz) to
$\nu_{K=0}$ (220.747 GHz), were all detected in IRS 1 (see Figure 7).
Similar to what is seen in OCS(19-18), the CH$_3$CN(12-11) lines
also show an emission peak at $\sim-59.5$ km s$^{-1}$
with weaker redshifted components.
At the intensity peak of IRS 1 SW,
the line profiles of $K$=0-8 transitions are shown in Figure 7a and b,
while as for IRS 1 NE, only $K$=0-7 transitions are detected above 
3-$\sigma$ level.
The $K$=0 and $K$=1 transitions, however, only have a frequency
difference of 4.25 MHz ($\sim$5.8 km s$^{-1}$). We used double-component
Gaussian models to fit the $K$=0, 1 line profiles for SW and NE respectively,
which give uncertainties of 10\% for SW and 15\% for NE in line fluxes.
Another issue is that the velocity separation between $K$=0 and 1 transitions
is comparable to the velocity difference of the emission peaks toward
SW and NE. With our sub-arcsec resolution, IRS 1 NE
and SW are only marginally separated, therefore the blending of
the main $K$=1 emission from IRS\,1 SW with the
redshifted $K$=0 emission from IRS\,1 NE is expected.
Due to the contamination caused by the blending, the transition of
$K$=0 toward IRS\,1 NE should be used with caution in the
following analysis. As Figure 7b shows, the $K$=0 and $K$=2
transitions from IRS 1 SW are quite weak toward IRS 1 NE. Assuming
that the main $K$=1 emission from IRS 1 SW has a similar intensity, we
estimate that about 15\% of the observed intensity of the redshifted
$K$=0 emission toward IRS 1 NE is actually contributed by the blending
from the main $K$=1 emission. Toward IRS 1 SW the
error of the main $K$=1 emission due to the blending from the redshifted
$K$=0 emission is negligible.

The intensity of the $K$=3 component
for both IRS\,1 NE and SW is abnormally low compared to the other
$K$ components under the assumption that the CH$_3$CN components are
in local thermal equilibrium (LTE), homogeneous and optically thin.
As shown in Figure 7a and b for the SW component, the peak intensity of
the $K$=0-4 transitions are nearly identical, indicating
that the low $K$ transitions of CH$_3$CN in the SW component are
optically thick. The optical depth of $K$=3 component is even
higher due to its doubled statistical weight. Thus, the optically thick
components must be excluded in the fitting for kinetic temperature
with the assumption of optically thin and LTE (see Section 4.5).

In addition, we noticed that the intensity of the $K$=6 component
toward IRS\,1 NE is higher than the other components including
$K$=3, 4, 5 and 7. This could be attributed to two causes. First, the $K$=6
component has a much higher column density than the components of
$K$=4 and 5 due to its doubled statistic weight, but it still remains
optically thin compared to the components of $K$=0, 1, 2 and 3.
Second, the intensity of the $K$=6 component toward IRS\,1 NE is
likely to be affected by the line blending with HNCO(10$_{1,9}$-9$_{1,8}$)
at rest frequency $\nu_0=220.5848$ GHz from SW and
CH$_3^{13}$CN(12$_3$-11$_3$) at $\nu_0=220.6000$ GHz from NE itself.
From the comparison of Figure 7b and 7d, it seems that these possible
line blending caused a non-zero baseline in the frequency range
from 220.58 to 220.61 GHz toward NE, which might contribute up to 20\% to the
measured peak intensity of $K$=6 component of CH$_3$CN(12-11).

The possible blending from weaker molecular lines
is also examined. We noticed that the line
frequencies of C$_2$H$_5$OH are close to the $K$=1, 3 and 4
transitions of CH$_3$CN(12-11) and a CH$_3$OCHO-E line close to
$K$=3 transition of CH$_3$CN(12-11). However, according to the JPL
catalog, the typical intensities of those lines are only a few
percent of the CH$_3$CN lines, and the contamination would produce no
significant effects.

We constructed line images for each $K$ transition (not shown in
this paper) and fitted the intensity peak positions for both
the IRS\,1 SW and NE components based on the integrated line flux
images of $K$=0-8. We also fitted the line profiles of each
$K$ transition to determine their peak intensities ($\Delta S_{\rm L}$),
centroid velocity ($V_{\rm LSR}$) and FWHM line width ($\Delta V_{\rm FWHM}$).
The derived values for these quantities are summarized in Table 4.

\begin{figure*}[t]
  \centering
  \begin{minipage}[]{0.6\linewidth}
  \includegraphics[width=9cm, angle=-90, origin=c]{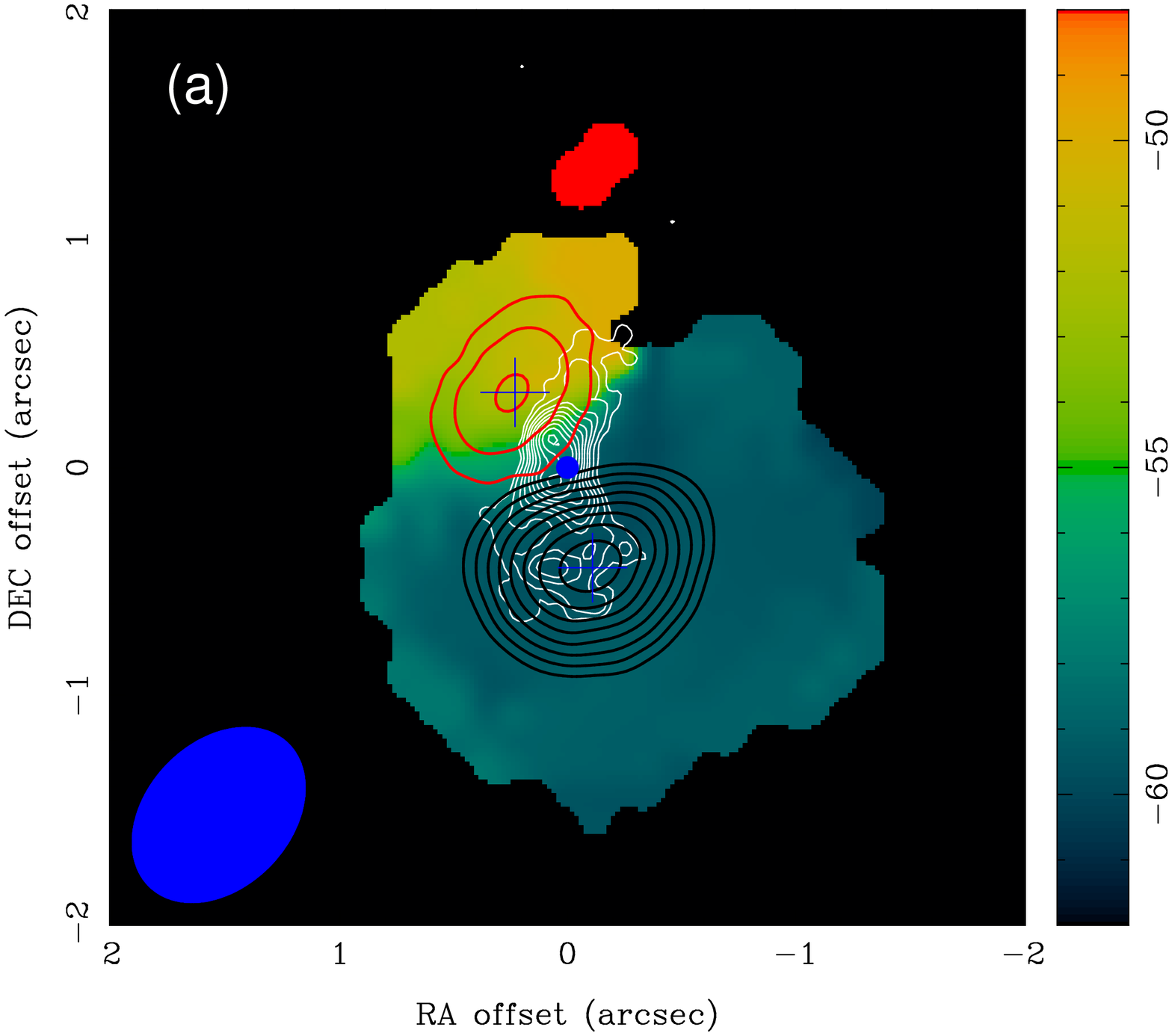}
  \end{minipage}
  \begin{minipage}[]{0.3\linewidth}
  \vspace{-1cm}
  \includegraphics[width=6.25cm, angle=0, origin=c]{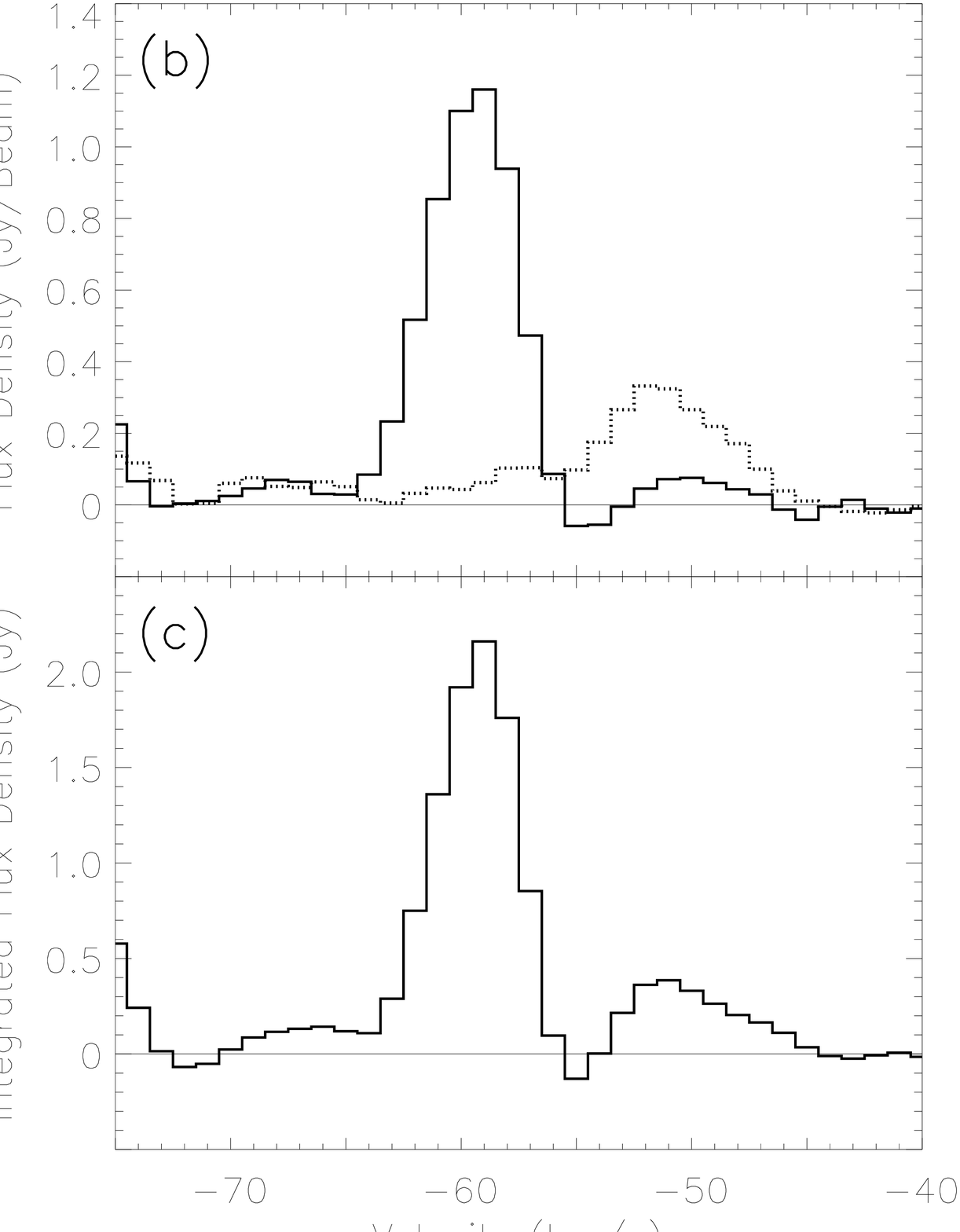}
  \end{minipage}
  \vspace{-0.5cm}
  \caption{(a) Same as the Figure 6a, but the red and black contours are
  for the combined CH$_3$CN(12-11) $K$=2-5 line. The velocity ranges
  for integrating ($-54.0$, $-47.0$) km s$^{-1}$ and ($-62.5$, $-57.0$)
  km s$^{-1}$ for IRS\,1 NE (red) and SW (black), respectively.
  The contour levels are (0.7, 1.1, 1.4) and (2.8, 3.2, 3.5, 3.9, 4.2, 4.6, 4.9)
  Jy beam$^{-1}$ km s$^{-1}$ for IRS\,1 NE and SW, respectively.
  (b) The line profiles of CH$_3$CN(12-11) $K$=2-5 lines at the
  positions of IRS 1 SW (solid line) and NE (dashed line). (c) The line profile of
  CH$_3$CN(12-11) $K$=2-5 lines integrated over the entire IRS\,1 region.
  }
\end{figure*}

Taking the strong transitions $K$=2-5, with no blending, we
combined the visibility data of the four transitions in the
LSR-velocity domain to construct the channel maps of CH$_3$CN(12-11)
(Figure 8), which is equivalent to averaging the line intensities
from the $K$=2-5 transitions. This spectral line image is
referred to as the $K$=2-5 line image, hereafter. Figure 9a shows
the integrated line intensity image of IRS 1 SW and NE
made from the $K$=2-5 line image, which is in good agreement with
that of OCS(19-18) (see Figure 6a). The peak position of IRS\,1 SW is
($-0\farcs11\pm0\farcs03$, $-0\farcs44\pm0\farcs03$) offset from
IRS\,1, based on a Gaussian fitting, while IRS\,1 NE is offset by
($0\farcs23\pm0\farcs07$, $0\farcs33\pm0\farcs09$), consistent with OCS(19-18).
The centroid velocity distribution made from the
CH$_3$CN(12-11) $K$=2-5 line image is also shown in Figure 9a, showing
a similar velocity gradient from SW to NE as that derived from the OCS(19-18)
line (see Figure 6a).

Figure 9b shows the spectral profiles made from the $K$=2-5 line image of
CH$_3$CN(12-11) toward the peak positions of IRS\,1 SW and NE while
Figure 9c is the spectrum integrated over the entire IRS\,1 region.
Gaussian fitting gives $V_{\rm LSR}=-59.5\pm0.1$ km s$^{-1}$ and
$\Delta V_{\rm FWHM} =4.3\pm0.1$ km s$^{-1}$ for IRS\,1 SW, and
$V_{\rm LSR}=-51.4\pm0.2$ km s$^{-1}$ and $\Delta V_{\rm FWHM}=6.8\pm0.5$
km s$^{-1}$ for IRS\,1 NE, respectively. These  results
are in good agreement with those from OCS(19-18) .

In addition, a weak (more than 3$\sigma$, 1$\sigma$=0.03 Jy beam$^{-1}$)
absorption of the CH$_3$CN(12-11) line at $V_{\rm LSR}=-56$ km s$^{-1}$ was
observed from the $K$=2-5 line image (see Figures 8 and 9c).
We will further discuss these possible absorption features.


\subsection{Kinetic Temperature of The Hot Molecular Gas}

As a symmetric top molecule, CH$_3$CN is a good probe for
determination of the kinetic temperature of the molecular
gas. The multiple $K$ transitions for $J\rightarrow J-1$
in a rotational transition of CH$_3$CN can be observed
simultaneously, and their line flux ratios reflect
the rotational temperature $T_{\rm rot}$, equivalent to
the kinetic temperature $T_{\rm k}$ under LTE condition.
Using rotational temperature equilibrium
analysis \citep[e.g.][]{holl82,lore84,chur92,gold99,aray05,furu08},
one can determine the kinetic (rotation) temperature $T_{\rm rot}$
and total column density $N$ of the hot molecular gas in an
optically thin case by fitting a linear function to the
relation between the (logarithmic) column density $N_{JK}$ and
the rotational energy $E_{JK}$ at levels ($J, K$) of CH$_3$CN:
\begin{equation} \label{} {\rm
ln}\left(\frac{N_{JK}}{g_{JK}}\right) = {\rm ln}\left[\frac{N}{Q(T_{\rm
rot})}\right]-{\rm ln}\left[\frac{\tau_{J,K}}{1-{\rm
e}^{-\tau_{J,K}}}\right]-\frac{E_{JK}}{kT_{\rm rot}},
\end{equation}
where $g_{JK}$ is the statistical weight of
the level $(J, K)$, $N$ is the total column density,
$\displaystyle Q(T_{\rm rot})=3.89T_{\rm
rot}^{1.5}/(1-{\rm exp}(-524.8/T_{\rm rot}))^2$ is the partition
function, and $k$ is Boltzmann constant.

In the case of NGC\,7538 IRS\,1, some of the $K$ components
of CH$_3$CN(12-11) are obviously optically thick (see Section 4.4).
Therefore, we made iterations in fitting the rotation temperature.
First, higher $K$ components
($K$=4, 5, 7 and 8 for SW, and $K$=4, 5, and 7 for NE) were assumed to
be optically thin and used in the initial fitting
with Equation (1) by eliminating the
optical-depth term $-{\rm ln}(\tau_{J,K}/(1-{\rm exp}(-\tau_{J,K})))$.
The initial fitting gave a rotation temperature $T'_{\rm rot}$ and a total
column density $N'$, with which the corresponding optical
depths $\tau'_{J,K}$ and optical-depth-correction factors
$f'_\tau(J,K)=\tau_{J,K}/(1-{\rm exp}(-\tau_{J,K}))$ were derived for each $K$
components. In the following iterations, we used $f'_\tau$ to correct
the observed surface column densities $N^{\rm obs}_{JK}$ and added the
other $K$ components to the fitting. To avoid the
bias caused by the optically thick $K$ components, their weighting
in the fitting was inversely proportional to their deviations
from the fitting curve of the previous iteration.

The final least-square fitting results are shown in Figure 10, giving
rotational temperatures of 260$\pm30$ K and 260$\pm$60 K for
IRS\,1 SW and NE, respectively. The derived rotational
temperatures are consistent with 297 K reported by \cite{klaa09}
and 245 K reported by \cite{qiu11}. 
We note that the optically thick components, $K$=0, 1, 2 and 3
for IRS\,1 SW and $K$=3 one for NE, significantly deviate from
the fitting curve, and have little influence on the
fitting results due to their low weighting. The derived
optical-depth-correction factors $f_\tau(J,K)$, which are based on
the fitting results ($N$ and $T_{\rm rot}$), are not adequate to
correct the $N_{JK}$ to the fitting curve. The optical depths of
the $K$ components used in the initial iterations of the fitting
are non negligible.

For the high $K$ components with smaller optical
depths, the correction factor $f_\tau$ is close to unity and not
sensitive to $\tau$, while for the optically thick components,
$f_\tau$ is almost proportional to $\tau$.
An underestimation of the $\tau_{J,K}$ of the high $K$ components
will not change the slope of the fit, which is largely
determined by these levels. However, the fitted total
column density will cause significant underestimations
for the $\tau_{J,K}$ and $f_\tau(J,K)$ of the optically thick levels,
which results in the large deviation from the fit (see Figure 10).

In summary, by fitting the $K$ components of
CH$_3$CN(12-11) lines with small optical depths, we
estimated the kinetic temperature of the hot
molecular clumps to be $\sim$260 K. The corresponding total column densities
for IRS\,1 SW and NE, however, might be underestimated by a factor of a few
and represent the lower limits.
With the derived $T_{\rm rot}$, we can estimate the lower limits of the total column densities
$N=1.9\pm0.5\times10^{16}$ cm$^{-2}$ and $N=3.5\pm1.5\times10^{15}$ cm$^{-2}$
toward the intensity peaks of SW and NE, respectively. With
an abundance of 1$\times$10$^{-8}$ for CH$_3$CN
\citep[][and references therein]{qiu11}, these column densities
correspond to gas masses of 4 and 0.7 M$_\odot$ for SW and NE
(in one FWHM beam size), respectively.

\begin{figure}[b]
  \centering
  \includegraphics[width=8cm,angle=0]{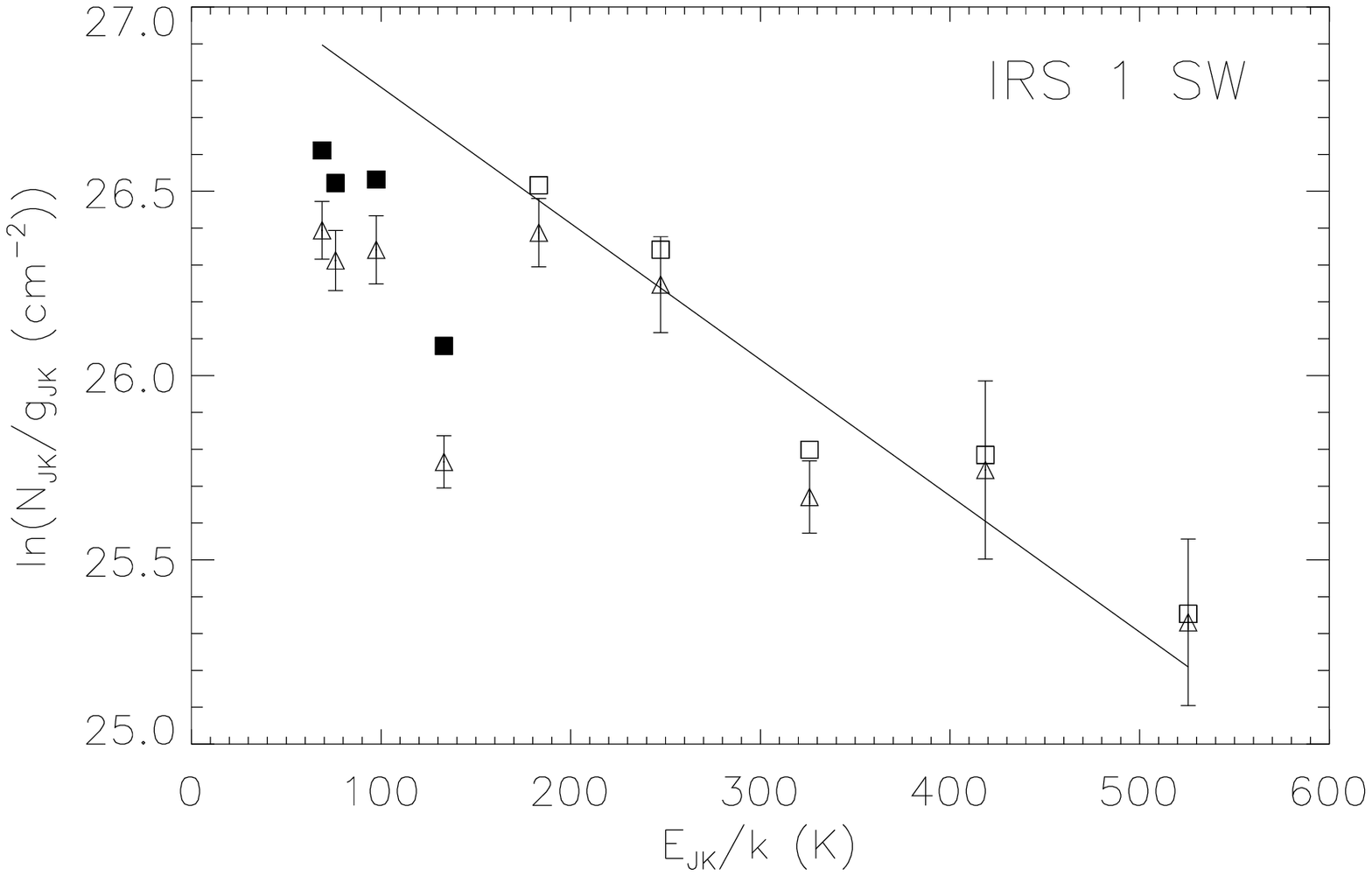}
  \vspace{-0.5cm}
  \\
  \includegraphics[width=8cm,angle=0]{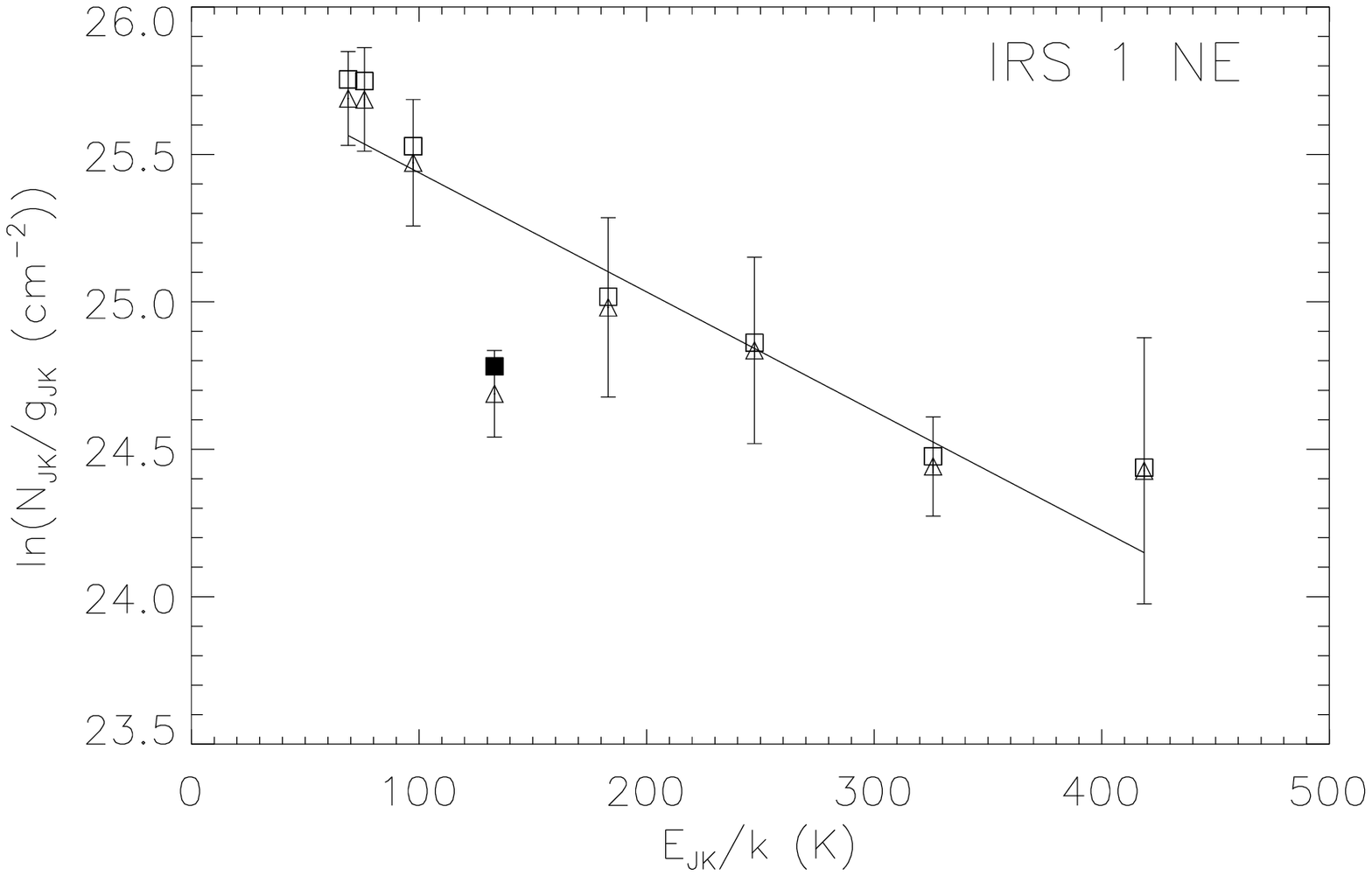}
  \caption{The population diagrams with $K$=0, 1, 2, 4, 5, 6, 7, 8
  transitions of CH$_3$CN(12-11) lines for IRS\,1 SW,
  and $K$=0, 1, 2, 4, 5, 6, 7 transitions for NE.
  The vertical bar marks 1$\sigma$ error. The corresponding
  measurement of $K$=3 transition is marked with filled square
  but excluded from the fitting.
  The straight lines are the best fittings to the data based on
  linear regression for ${\rm  ln}(\frac{N_{JK}}{g_{JK}})
  =\alpha \frac{E_{JK}}{k} +\beta$, with $\alpha=-0.0038\pm0.0004$
  and $\beta=27.5\pm0.1$ for IRS\,1 SW and $\alpha=-0.0038\pm0.0012$
  and $\beta=25.8\pm0.4$ for NE.
  }
\end{figure}


\subsection{$^{13}$CO(2-1), CO(2-1) \& HCN(1-0) toward IRS\,1}

\subsubsection{$^{13}$CO(2-1) toward IRS\,1}

The $^{13}$CO(2-1) line is a good tracer for the gas with a
low to intermediate density. We produced a high-velocity
($\delta V=0.5$ km s$^{-1}$) and high-angular (0\farcs8$\times$0\farcs6)
resolution image of the $^{13}$CO(2-1) line to study both
the absorption line against the continuum source and the emission
line from the molecular gas immediately surrounding IRS\,1.
Figure 11 shows the channel maps of $^{13}$CO(2-1), exhibiting
a strong absorption between $-60$ to $-53$ km s$^{-1}$ toward
the 1.3-mm continuum peak and an emission component SW to IRS 1
in the velocity range of ($-63$, $-60$) km s$^{-1}$. A similar
distribution of $^{13}$CO(2-1) gas is also shown in Figure 12,
the integrated intensity maps of both $^{13}$CO(2-1) emission
and absorption from the velocity range between $-43$ to $-66$ km s$^{-1}$.
The morphology of the
integrated $^{13}$CO(2-1) emission shows an elongated feature
across IRS\,1 with a P.A.$\sim$35\arcdeg. At an offset
$\sim$1\arcsec\, southwest of IRS\,1, a blueshifted gas
component is also observed from $^{13}$CO(2-1) emission.
In general, both the integrated line flux and the
intensity-weighted velocity images derived from the $^{13}$CO(2-1)
line appear to be in good agreement with the results observed
from the molecular lines OCS(19-18) and CH$_3$CN(12-11).

\begin{figure*}[]
  \centering
  \includegraphics[width=6cm, angle=-90]{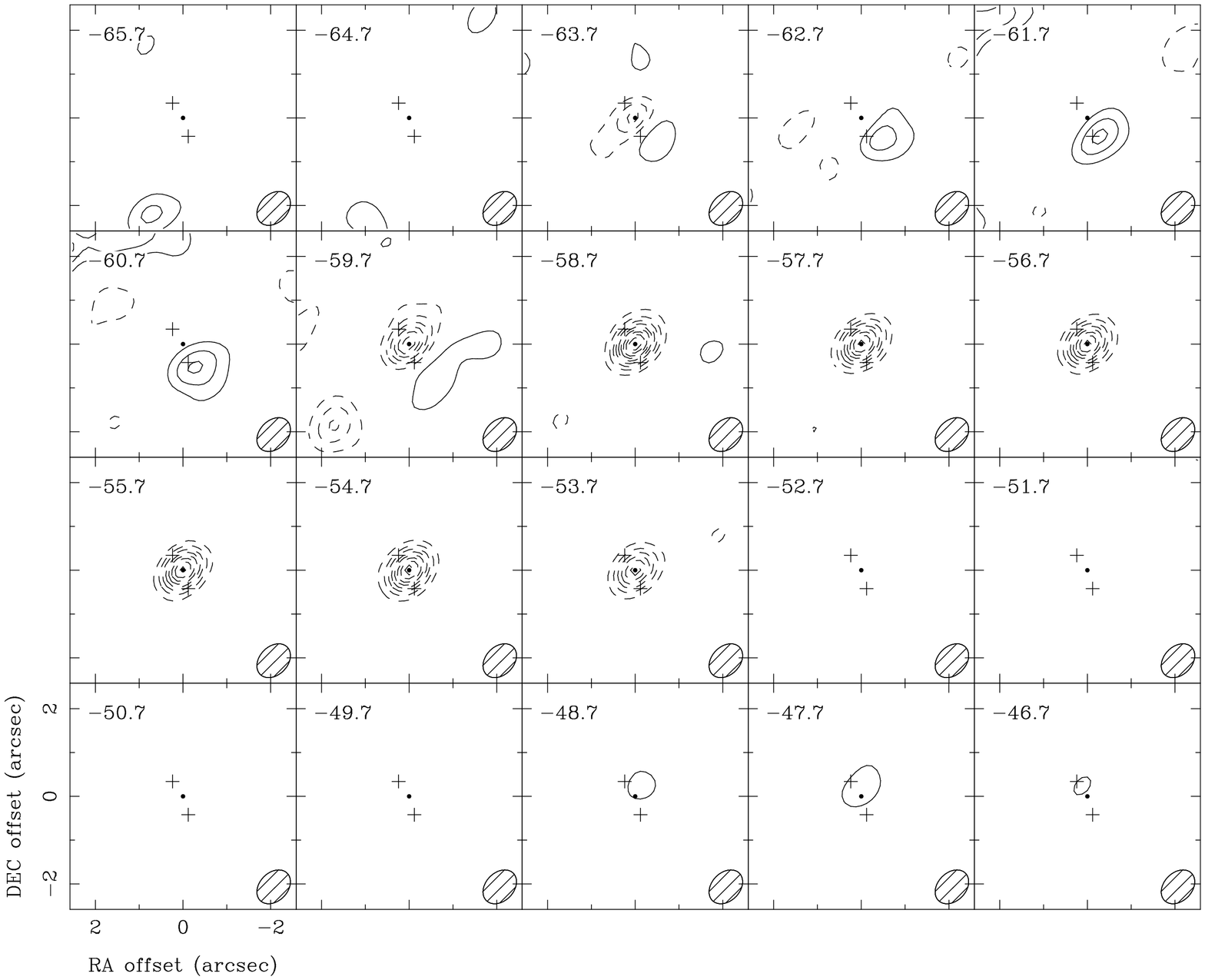}
  \caption{The channel maps of $^{13}$CO(2-1).
  The contours are (..., --4, --3, --2, --1, 1, 2, ...)$\times$5$\sigma$,
  and 1 $\sigma=0.05$ Jy beam$^{-1}$. The FWHM beam size is
  0\farcs8$\times$0\farcs6, $\rm P.A.=-44$\arcdeg. IRS\,1 is located
  at the phase center (indicated by the filled black circle).}
\end{figure*}

\begin{figure*}[]
  \centering
  \includegraphics[width=6cm, angle=-90]{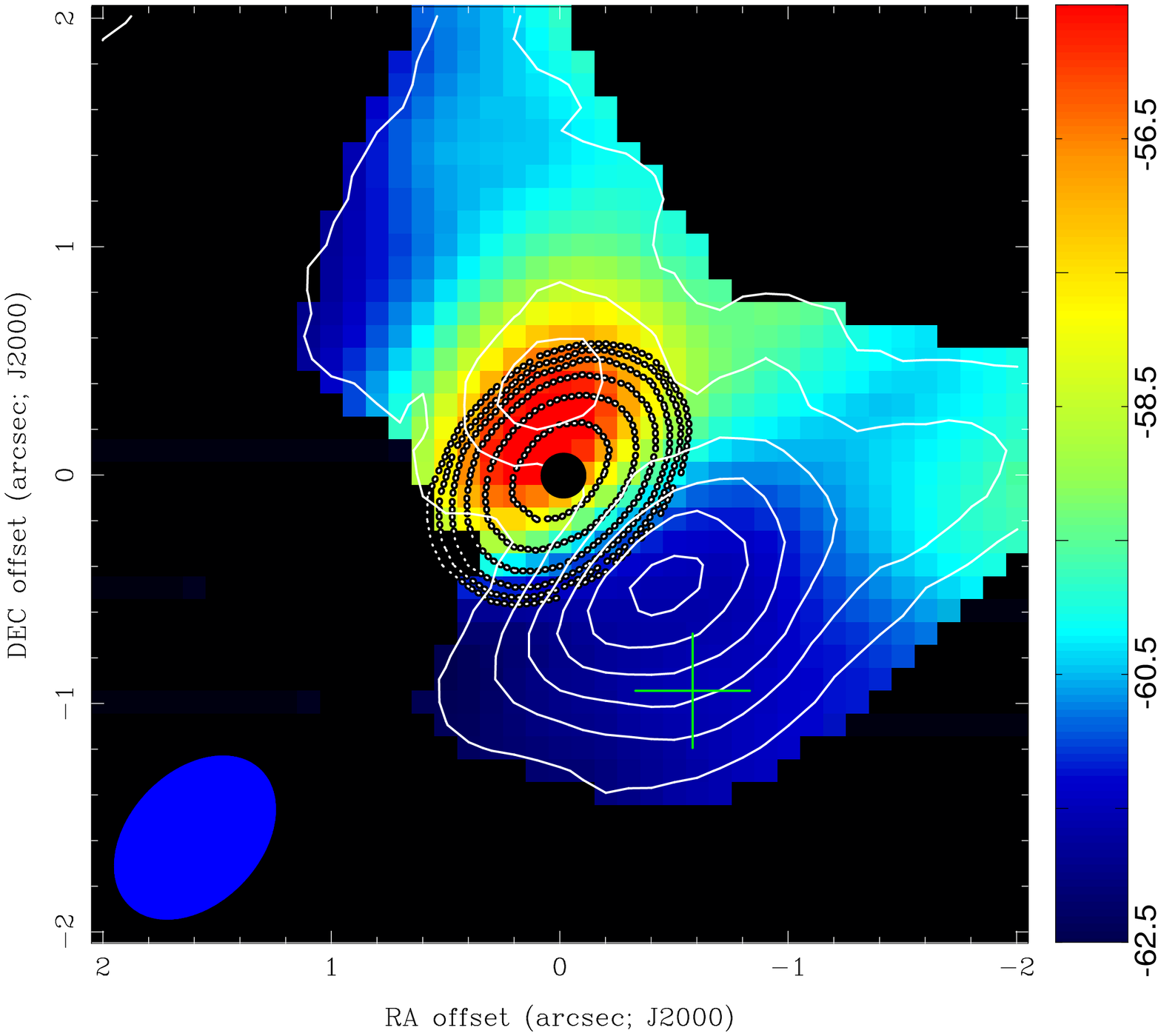}
  \caption{The images of $^{13}$CO(2-1) emission (solid
  white contours) and absorption (dotted contours) constructed
  by integrating the line intensity from the velocity range
  between $-43$ to $-66$ km~s$^{-1}$ with an FWHM beam size
  0$\farcs$8$\times$0$\farcs$6 (P.A. $=-45\arcdeg$). The emission contours
  are 2, 2.5, 3, 3.5, 4 and 4.5 Jy beam$^{-1}$~km~s$^{-1}$.
  The absorption contours are --2, --2.8, --4, --5.7, --8,
  --11.3 Jy beam$^{-1}$~km~s$^{-1}$. The intensity-weighted
  radial velocity (LSR) image (color) was made from the
  $^{13}$CO(2-1) emission line in the same velocity range
  used for the total intensity image of the line emission. The
  color wedge scales the velocities in the units of km s$^{-1}$.
  The green plus sign indicates the emission peak of the SW 1.3-mm
  continuum extension.
  }
\end{figure*}

We made a $^{13}$CO(2-1) spectrum toward the 1.3-mm emission peak
in IRS\,1 region and fitted the line profiles with a
multiple-Gaussian-component model (see Figure 13a).
A narrow ($\Delta V_{\rm FWHM}\sim$
1 km s$^{-1}$) and strong ($\Delta S_{\rm L}=-2.6$ Jy beam$^{-1}$)
absorption feature at $-59.0$ km s$^{-1}$ was observed toward IRS\,1,
which is best seen in a spectrum made from using the long ($>$100
k$\lambda$) baseline data. The narrow velocity characteristics
of this feature suggest that a large fraction of the absorption
at the velocity close to the systemic velocity arises from the
absorption by the cold gas in the molecular envelope located in
front of the continuum source IRS\,1. The absorption peak velocity
in $^{13}$CO(2-1)  verifies that the systemic velocity of IRS
1 is very close (within 0.5 km s$^{-1}$) to $V_{\rm sys}=-59.0$ km
s$^{-1}$ as observed in other molecular lines.
Given $\Delta S_{\rm L}=-2.6$ Jy beam$^{-1}$ and
$S_{\rm C}=3.25$ Jy beam$^{-1}$
the peak optical depth of the $^{13}$CO(2-1) line can
be determined for the main absorption feature near
the systemic velocity using Equation (1) of \citet{qin08},
$\tau_{\rm L} = - {\rm ln}\left(1 + \Delta
S_{\rm L}/S_{\rm C}\right) = 1.6$.

Two more absorption components are seen at $-57.0$ and $-54.0$ km
s$^{-1}$, which are likely due to infall toward the central source
in IRS\,1. For these two absorption features, the equivalent total
line width is $\sim$5 km s$^{-1}$, which is consistent with the
main absorption feature in the inverse P-Cygni profile suggested
from the lower-angular resolution observations of the HCO$^+$(1-0)
line \citep{cord08,sand09}. A weak (0.5 Jy beam$^{-1}$), and broad
($\Delta V_{\rm FWHM}=5.5$ km s$^{-1}$) emission at velocity
$-63.0$ km s$^{-1}$ was also detected, corresponding to the
blueshifted emission in the inverse P-Cygni profile from
lower angular resolution observations.

Two additional significant spectral features in $^{13}$CO(2-1) were
also detected: a blueshifted absorption component
($\Delta S_{\rm L}=-0.5$ Jy beam$^{-1}$, $\Delta V_{\rm FWHM}=1.3$ km s$^{-1}$)
at $V_{\rm LSR}=-63.8$ km s$^{-1}$ and a redshifted emission component
($\Delta S_{\rm L}=0.45$ Jy beam$^{-1}$, $\Delta V_{\rm FWHM}=2.5$ km s$^{-1}$)
at $V_{\rm LSR}=-48.0$ km s$^{-1}$.
The blueshifted absorption component is probably due to
the foreground outflow gas, for which we find
$\tau_{\rm L} = - {\rm ln}\left(1 + \Delta S_{\rm L}/S_{\rm C}\right) \approx0.17$.
While the redshifted emission at $V_{\rm LSR}=-48.0$ km s$^{-1}$
could also be due to outflow,
it more likely traces a background cloud emission at the same
radial velocity, as indicated by lower-resolution studies on the
general gas distribution in the NGC 7538 region (Sandell 2013,
private communications).

To estimate the column density of the observed $^{13}$CO gas, a reasonable
excitation temperature $T_{\rm ex}$ is needed. The absorption
feature of $^{13}$CO (2-1) is seen only in the central 1\farcs5 region
in projection. If the $^{13}$CO gas is all within the central 1\farcs5,
and co-exists with the hot molecular gas, then a high $T_{\rm ex}$ of
$\sim$200 K is appropriate. However, the lack of the $^{13}$CO absorption
outside the 1\farcs5 region might be simply due to the absence of
an emission background. In this case, the $^{13}$CO absorption arises from
gas along the line of sight, and a lower excitation temperature should be
considered. \cite{sand04} derived a $T_{\rm d}$ of 75 K from fitting
the dust emission in a $\sim$11\farcs5$\times$10\farcs6 region around IRS 1,
constrained by the sub-millimeter luminosity of IRS 1. We calculated
the total $^{13}$CO column density with both the high $T_{\rm ex}=200$ K
(upper limit) and low $T_{\rm ex}=75$ K (lower limit) for each velocity
component of $^{13}$CO (2-1) and listed the results in Table 5.

\begin{figure}[]
  \centering
  \includegraphics[width=4.25 cm, angle=0]{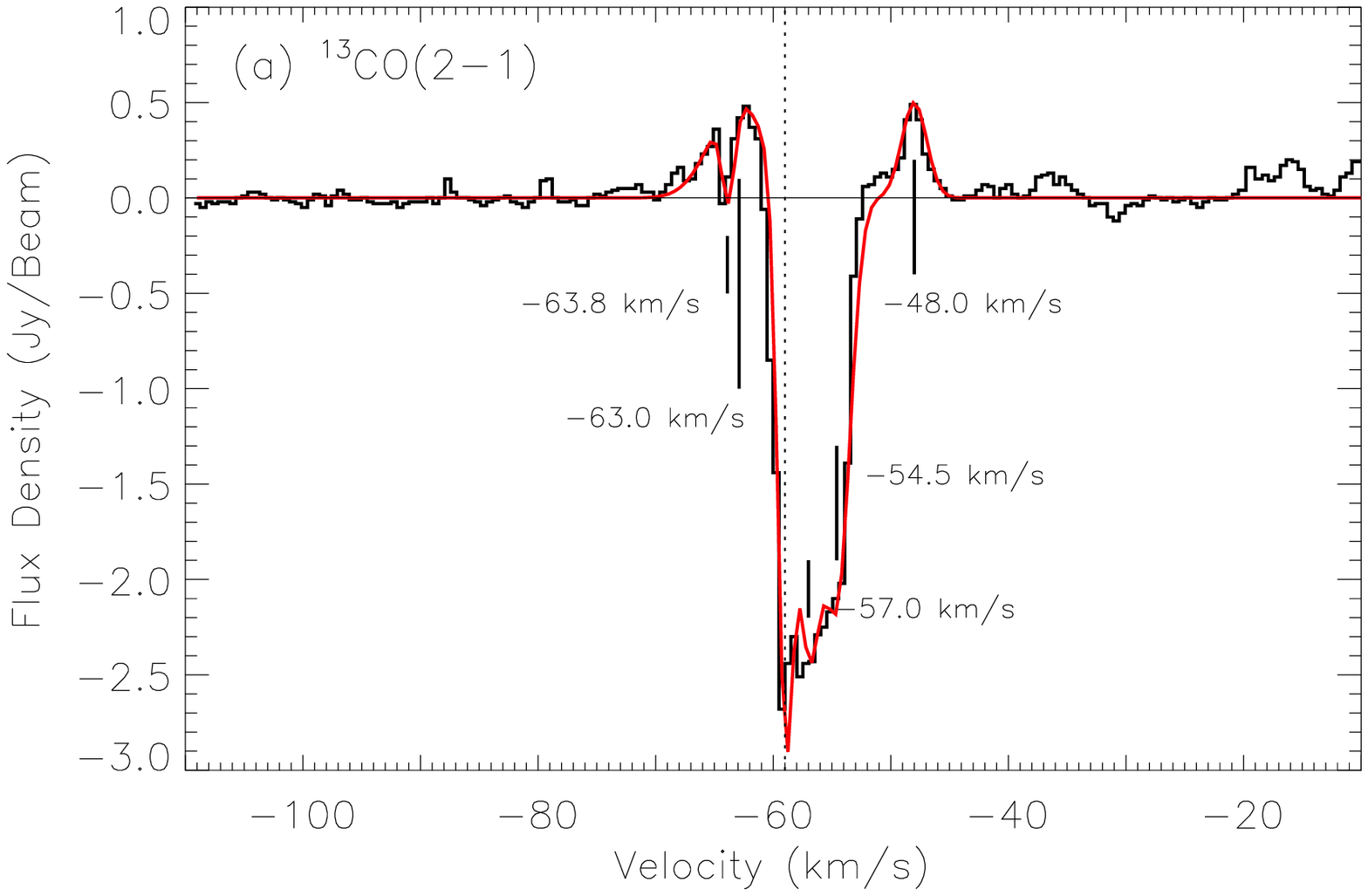}
  \hspace{-0.5cm}
  \includegraphics[width=4.25 cm, angle=0]{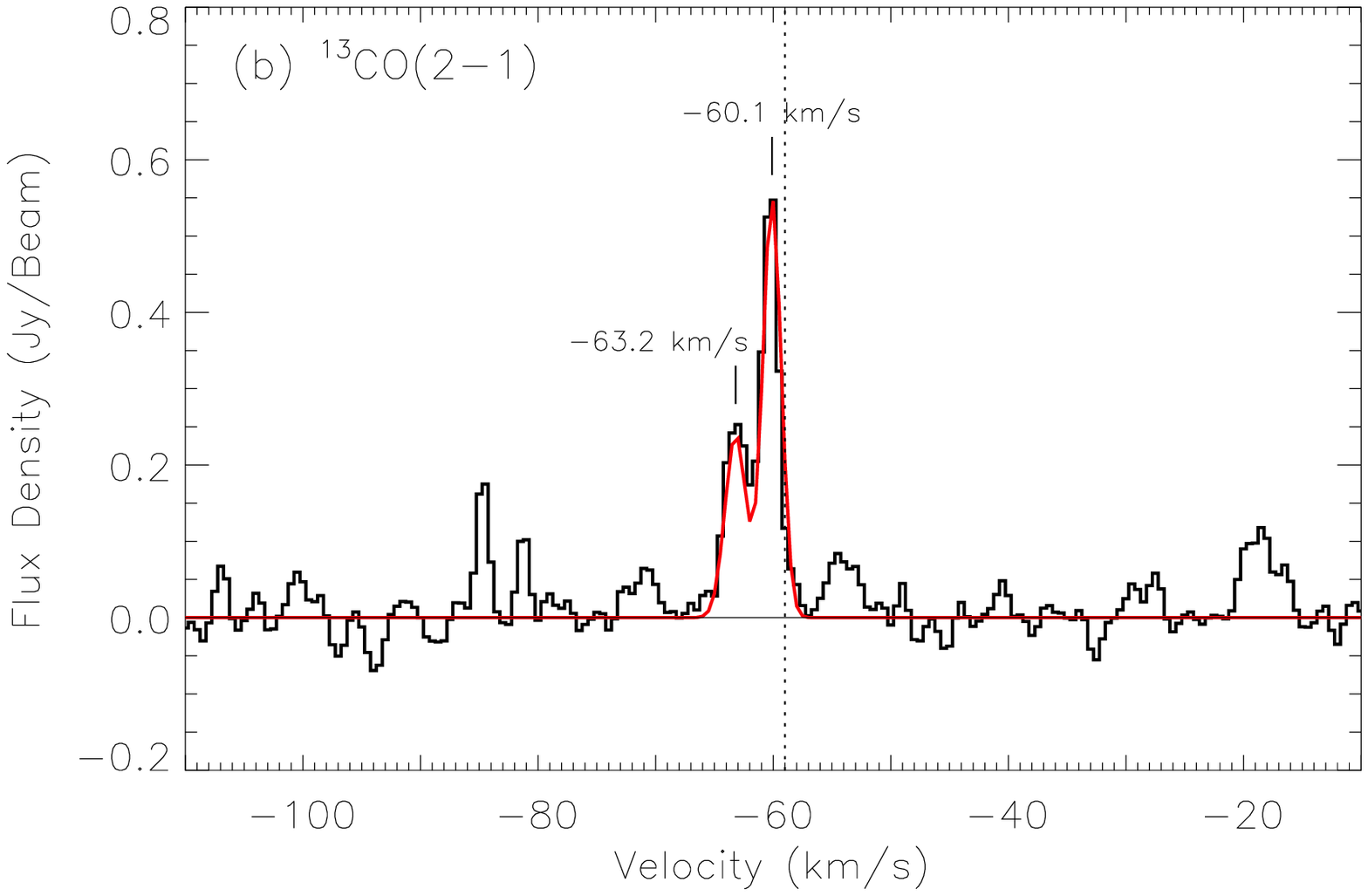}
  \\
  \includegraphics[width=4.25 cm, angle=0]{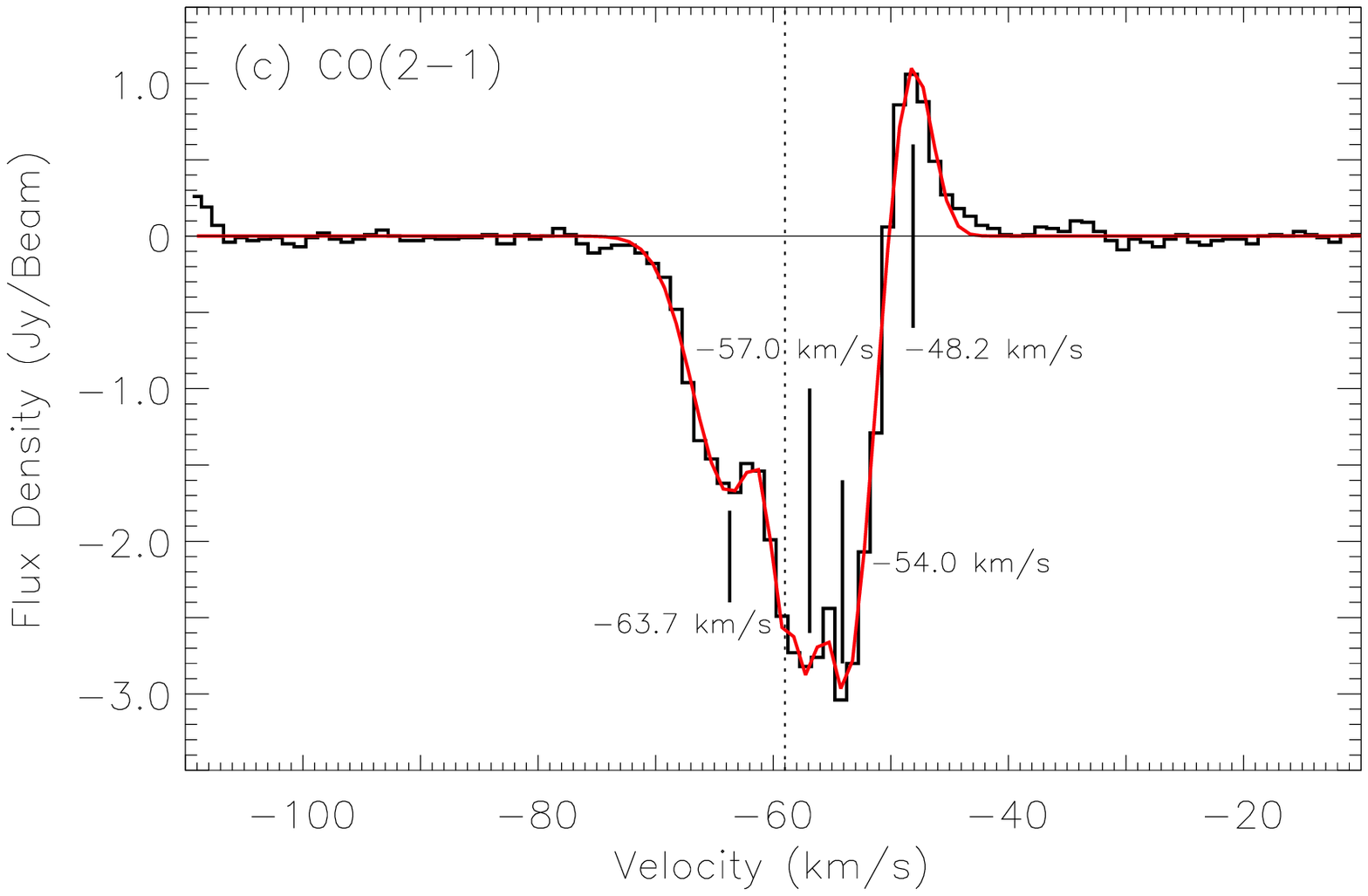}
  \hspace{-0.5cm}
  \includegraphics[width=4.25 cm, angle=0]{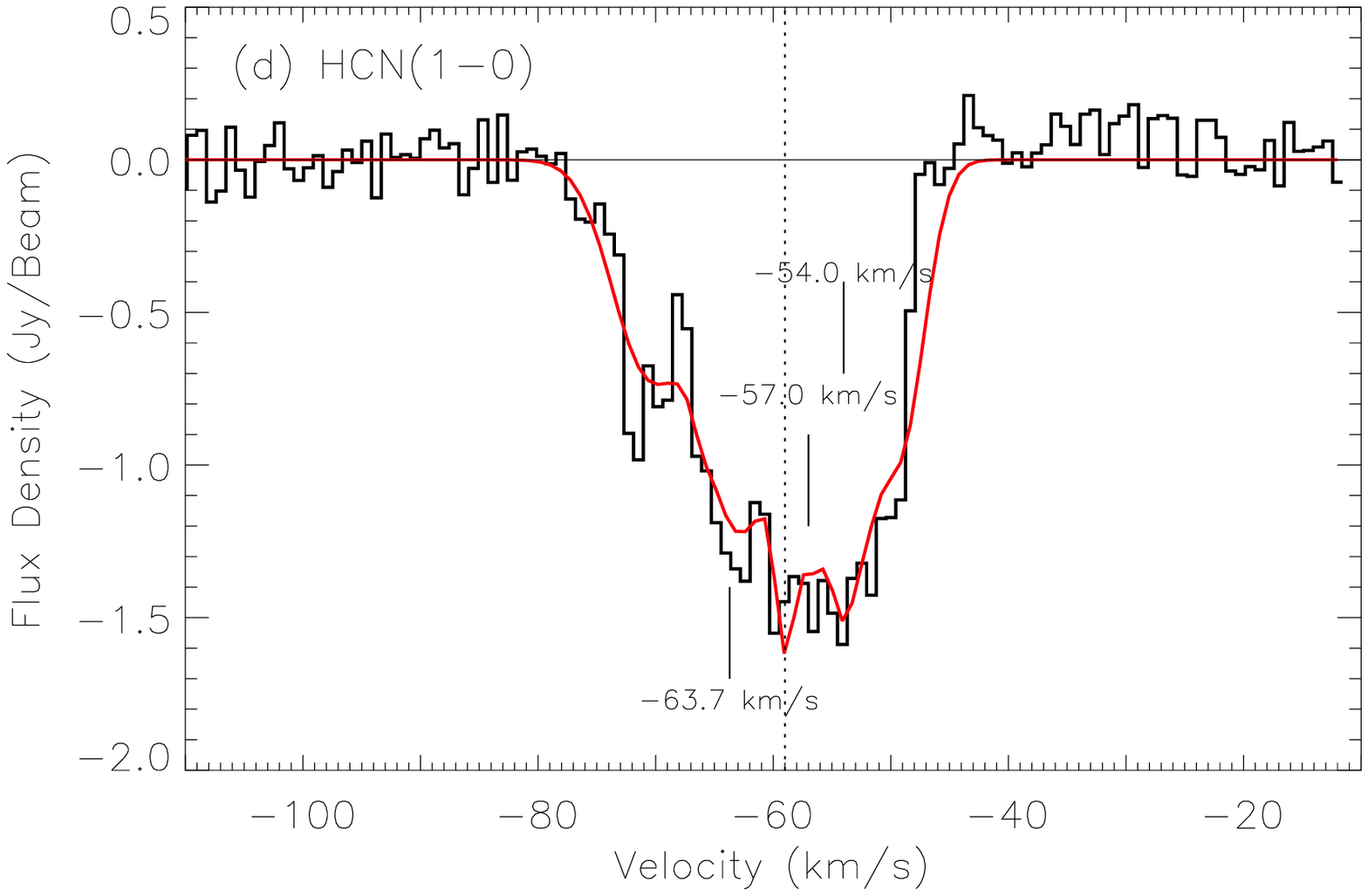}
  \caption{(a), (c) and (d): the $^{13}$CO(2-1), CO(2-1)
  and HCN(1-0) spectra toward IRS\,1, respectively.
  (b) the $^{13}$CO(2-1) spectrum toward the position of
  the 1.3-mm SW extension.
  In all the figures, the dashed vertical lines mark
  V$_{\rm sys}=-59.0$ km~s$^{-1}$, and the red curves are
  the multiple-Gaussian fitting to the line profiles.
  The parameters for each of the kinematic components
  from the fitting are summarized in Table 5.}
\end{figure}

\begin{figure*}[]
  \centering
  \includegraphics[width=6 cm, angle=-90]{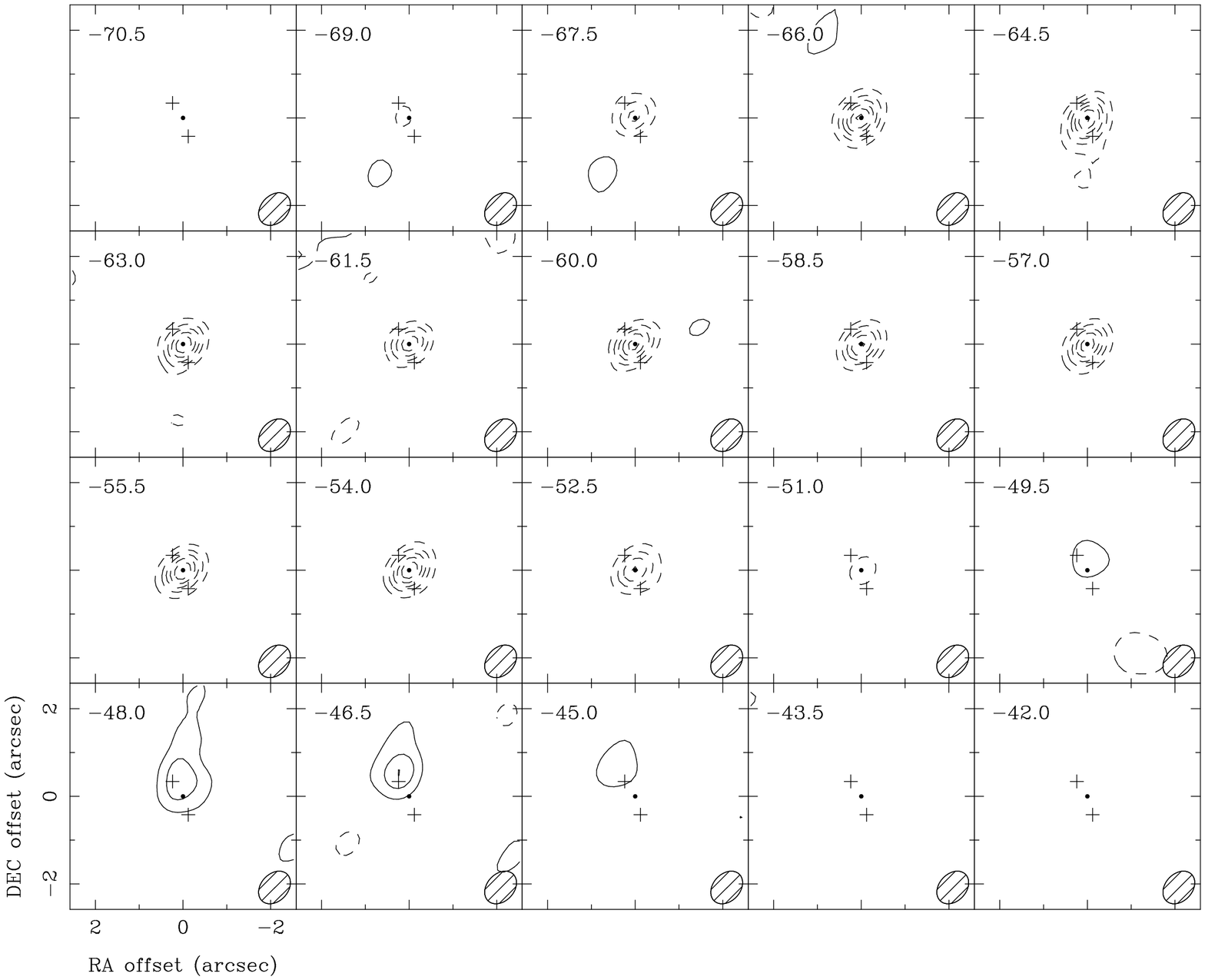}
  \caption{The channel maps of CO(2-1).
  The contours are (..., --4, --3, --2, --1, 1, 2, 3, 4...)$\times$5$\sigma$,
  and 1 $\sigma=0.1$ Jy beam$^{-1}$. The FWHM beam size is
  0\farcs8$\times$0\farcs6, $\rm P.A.=-42$\arcdeg. IRS\,1 is located
  at the phase center (indicated by the filled black circle).}
\end{figure*}

We also extracted a $^{13}$CO(2-1) spectrum toward the ``SW tail"
structure observed in the 1.3-mm continuum (Figure 13b).
Toward the position of the SW extension in
1.3-mm continuum, a narrow
($\Delta V_{\rm FWHM}\approx 1.8$ km s$^{-1}$) emission feature
at $-60.1$ km s$^{-1}$ with $\Delta S_{\rm L}=0.6$ Jy beam$^{-1}$
is seen close to
$V_{\rm sys}$, which seems to be affected
by the contamination from the spectrum from IRS\,1 and
has a skewed line profile. In addition, a narrow, weaker emission
feature ($\Delta V_{\rm FWHM}\sim 2.1$ km s$^{-1}$,
$\Delta S_{\rm L}=0.2$ Jy beam$^{-1}$) at $-63.2$ km s$^{-1}$
is detected which could be associated with low-velocity outflow
gas.

\subsubsection{CO(2-1) toward IRS\,1}

We made channel maps of CO(2-1) (Figure 14), showing
a wide absorption toward IRS\,1 from $-69$ to $-52$
km s$^{-1}$. Multiple spectral components are included
in this wide absorption, which are best seen in
the CO(2-1) spectra toward the 1.3-mm emission
peak in IRS\,1 (Figure 13c) and the Gaussian fitting.
Absorption features
at velocities $-63.7$, $-59.0$, $-57.0$ and $-54.0$ km
s$^{-1}$, an emission feature at $-48.2$ km
s$^{-1}$, and the fitting results are summarized in
Table 5. The fitting results are in good agreement with
 $^{13}$CO(2-1) except for the absorption
feature at $-63.7$ km s$^{-1}$ which is much deeper
($\Delta S_{\rm L}=-1.7$ Jy) and much broader
($\Delta V_{\rm FWHM}=7$ km s$^{-1}$) than its
$^{13}$CO(2-1) counterpart. The CO(2-1) counterpart
corresponding to the blueshifted emission observed in
$^{13}$CO(2-1) at $-63.0$ km s$^{-1}$ was not detected,
which might be diluted by the strong and broad absorption
feature at $-63.7$ km s$^{-1}$. The wide blueshifted
CO(2-1) absorption is likely due to foreground
outflow gas. The optical depths of the
CO(2-1) features are determined with
$\tau_{\rm L}= -{\rm ln}(1-\Delta S_{\rm L}/S_{\rm C})$,
showing the large optical depths ($\sim 3$) of the CO(2-1)
components that are associated with the redshifted
absorption gas, and suggesting that the infalling CO(2-1) gas
is optically thick.

\subsubsection{HCN(1-0) Hyperfine Lines \& Absorption Profile}

The $J$=1-0 transition of HCN has three hyperfine lines
($F$=0-1, 1-1 and 2-1) with optical depths in the ratio 1:3:5 under LTE.
The strongest line $F$=2-1 has $\nu_0=88.6318$ GHz.
The velocity separations of $F$=1-1 and $F$=0-1 with
respect to $F$=2-1 are +4.84 and $-$7.08 km s$^{-1}$,
respectively. We made multiple Gaussian
fits to the four kinematic components in absorption
identified in CO(2-1) to the HCN(1-0) absorption profile (Figure 13d).
For each kinematic component, we used the three hyperfine
lines of the HCN(1-0) transition in the initial fitting.
In the fits, the intensity ratio of each hyperfine line was
fixed assuming LTE, and the FWHM line widths for the hyperfine
components were kept to be the same for each kinematic component.
The three parameters $\Delta S_{\rm L}$, $V_{\rm LSR}$ and
$\Delta V_{\rm FWHM}$ for the main hyperfine lines ($F$=2-1)
are determined from the best fitting. However, the hyperfine
lines $F$=0-1 and $F$=1-1 in the high blueshifted and high
redshifted ends of the absorption profile require high
intensities that do not fit to their corresponding LTE ratio.
The high-velocity HCN line features might be non-LTE.
Alternatively, the broadening of the HCN(1-0) absorption profile
might be caused by high-velocity gas components which are associated
with shocked or compressed gas and traced by the HCN emission \citep[e.g.][]{shi10}.
The absorptions caused by these high-velocity gas components might be
insignificant in CO(2-1) due to the mitigation of the emission of the CO(2-1) outflow.
Table 5 summarizes the fitting results.


\section{Discussion}

\subsection{A Massive Ionizing Star in the HC \ion{H}{2} region IRS\,1}

IRS\,1 is a HC \ion{H}{2} region associated with a newborn
star. At the high-angular resolution ($\sim$0\farcs3) in the
CARMA observations, the surrounding dust emission has been
resolved out. Fitting the continuum fluxes of IRS 1
at multiple wavelengths indicates that even at 1.3-mm the free-free
emission still dominates the continuum emission \citep{sand09}.
If we assume optically thin emission and neglect
the contribution of the thermal dust emission, a Lyman
continuum flux $N_{\rm Lym}$ can be estimated from the 1.3 mm flux
density ($S_{\rm C}=3.8$ Jy at $\nu=224$ GHz) of IRS\,1 with the
equation below of a recombination coefficient of
$\alpha_{\rm B}=2.58\times10^{-13}$ cm$^{-3}$ s$^{-1}$ in case B and
electron temperature of $T_{\rm e}=1\times10^4$ K \citep{humm63},
\begin{eqnarray}
N_{\rm Lym}&=&7.6\times10^{46}\, {\rm phot\, s^{-1}} \, \alpha^{-1}(\nu, T_{\rm e})
\times\nonumber \\
           & & \left(\frac{T_{\rm e}}{10^4\,K}\right)^{0.35}
                                          \left(\frac{\nu}{\rm GHz}\right)^{0.1}
                                          \left(\frac{S_{\rm C}}{\rm Jy}\right)
                                           \left(\frac{D}{\rm kpc}\right)^2,
\end{eqnarray}
where the correction factor for power-law approximation
$\alpha(\nu, T_{\rm e}) = 0.85$ for $T_{\rm e}=10,000$ K
and $\nu=224$ GHz \citep{mezger67}. For $D=2.65$ kpc, we
have $N_{\rm Lym}=4.1\times10^{48}$ phot s$^{-1}$.
At 224 GHz, our observations might not be sensitive
enough to reveal the extended structures of the ionized outflow
as the VLA observations shown (see Figure 2).
However, with Equation (2) and the U-band measurement
\citep[60 mJy, see][]{sand09}, it suggests that only
$\sim$2$\times$10$^{46}$ photo s$^{-1}$ was contributed by
the extended structure, less than 1\% of the 224-GHz results.

However, a few caveats need to be kept in mind for
the above estimate. First, most of the free-free emission
in the unresolved central region ($\le$0\farcs1) is actually
optically thick, resulting in underestimation of the uv photon flux
from the YSO in IRS 1. Second, the contribution of
dust thermal emission might not be negligible. If we extrapolate the total
flux density fit for IRS 1 from the spectrum between 4.8 and 43.4 GHz
\citep{sand09} to 224 GHz, a total flux of $\sim$2.6 Jy from
free-free emission is predicted, implying that dust thermal
emission might contribute $\sim$30 percent of the total flux density. 
In addition, ionization is not only due to
the UV photons from the stellar photosphere but also caused by the
heating in the accretion process.
For low-mass pre-main-sequence stars (classical T Tauri stars, CTTSs),
the shock front due to accretion will produce hot plasma with a
temperature of a few times of 10$^6$ K and yield significant emission
in the soft X-ray band \citep{argi12}. It is possible that in the case of
high-mass YSOs with ongoing accretion similar processes also occur,
and produce high-energy photons which will ionize surrounding molecular
gas. Unfortunately, currently there is no detailed quantitative model on how
much Lyman continuum flux the accretion process can contribute for the
high-mass YSOs. However, we believe that the uv photons from the stellar
photospheres of the YSOs are the dominating source for photoionization.

Neglecting all the factors described above,
this Lyman continuum flux of $\sim$4.1$\times$10$^{48}$ phot s$^{-1}$
is equivalent to the flux from a single star of
O8-O9.5 in the main sequence \citep{vacc96,mart05}, which is required to
maintain the ionization of the HC \ion{H}{2} region IRS\,1 alone.
Taking account of the possible overestimation of 30\% due to dust thermal
emission will not change the derived spectral type.
Considering the fact that we might significantly underestimated the continuum flux
due to optical depth effects, the assessment for the spectral type of the
ionizing star is also consistent with the suggestions of an O6/7 star
estimated previously \citep[e.g.][]{will76,akab05} if the turn-over frequency for
the free-free emission from the HC HII region is significantly higher than
230 GHz ($\lambda\sim$1.3 mm). Then, the
massive ionizing star should be the primary energy source, responsible for
ionizing the HC \ion{H}{2} region and driving the energetic ionized outflow
via the overwhelming radiative pressure.

\subsection{Infalling Gas}

As suggested by \cite{cord08}, \cite{sand09} and \cite{beut12},
an on-going accretion might be taking place in the IRS\,1 system.
From the $^{13}$CO(2-1) absorption in front of the IRS\,1 core,
the total column density of molecular gas is 2.0$\times10^{24}$
cm$^{-2}$ using [CO]/[H$_2$]=10$^{-4}$, [$^{13}$CO]/[CO]=1/60
and a high $T_{\rm ex}$=200 K (see Table 5). If the molecular gas
is spherically and evenly distributed within a radius of 1900 AU
(equivalent to the 0\farcs7 FWHM beam size of the CARMA 1.3-mm
observations), this column density implies that about 5 M$_\odot$ in the
surrounding region is available for accretion onto IRS\,1. On the
other hand, if we adopt a large-scale distribution of the
$^{13}$CO gas (the lower column density of 8$\times$10$^{23}$ cm$^{-2}$
corresponding to the lower $T_{\rm ex}=75$ K needs to be considered),
the mass of the molecular gas associated with IRS\,1 would be
larger since its value is proportional to the square of radius.

The inverse P-Cygni profile revealed from the observations of
HCO$^+$(1-0) by \cite{cord08} \& \cite{sand09} suggests that
the infall likely occurs within the beam size of 4\farcs5, or
the inner region with a size of 0.06 pc. Our sub-arcsec
resolution observations of HCN(1-0), CO(2-1) and $^{13}$CO(2-1) lines toward
IRS\,1 agree with the previous results.

Absorption features are also detected in
OCS(19-18) and $K$=2-5 components of CH$_3$CN(12-11) 
(see Sections 4.3 and 4.4) in the velocity range bettwen $-56.5$ to
$-53.0$ km~s$^{-1}$, corresponding to the absorption features at
$-57.0$ and $-54.0$ km~s$^{-1}$ in HCN(1-0), CO(2-1) and $^{13}$CO(2-1).
These absorptions probably arise in
the infalling gas, revealing the hot dense gas in the accretion flow.
The hot molecular gas is more redshifted than the
corresponding HCN(1-0) and CO absorptions, perhaps tracing an accelerating
infall closer to IRS\,1.

The redshifted $^{13}$CO(2-1)
absorption against the continuum radiation from IRS\,1 clearly
shows evidence for ongoing gas infall. From the optical depth of
the infalling gas determined from the relatively optically thin line
$^{13}$CO(2-1), as well as the excitation temperature determined
from the CH$_3$CN population diagram, we can estimate the
infall rate $\displaystyle dM/dt=4\pi r^2 m_{\rm H} \mu n_{\rm H}
V_{\rm in}$, at an infall radius $r$, where the mean molecular
weight $\mu=2.35$ and density of the accretion materials
$n_{\rm H}= N_{\rm H}/r$.
Thus, the infall rate is a function of the measurable quantities,
\begin{eqnarray}
\frac{dM}{dt}&=& 2.8\times10^{-4} \, {\rm M}_\odot\, {\rm yr}^{-1}
                \left[\frac{\theta_{\rm in}}{\rm arcsec}\right]
                \left[\frac{D}{\rm kpc}\right] \nonumber \\
             && \left[\frac{N_{\rm H}}{\rm 10^{24}\,{\rm cm}^{-2}}\right]
                \left[\frac{\Delta V_{\rm in}}{\rm km~s^{-1}}\right],
\end{eqnarray}
where the angular size of the infall region $\theta_{\rm in}$
is approximately of the geometric mean of the telescope
beam $\theta_{\rm B}=0\farcs7$.

The two redshifted absorption components at $-54.5$ and $-57.0$
km s$^{-1}$ in $^{13}$CO(2-1) give H$_2$ column densities
($N_{\rm H,red}$) of 9$\times$10$^{23}$ and 1$\times$10$^{24}$ cm$^{-2}$,
respectively. If we take the velocity displacements of
the two absorption features from the systemic velocity
as their corresponding infall velocity
($V_{\rm in}\approx\Delta V=V_{\rm LSR}-V_{\rm sys}$),
an infall rate of 3$\times$10$^{-3}$ $\rm M_\odot$ yr$^{-1}$ can be
inferred based on equation (3) for the IRS\,1 region,
higher than the value derived by previous study \citep{cord08,sand09},
lower than that of \cite{beut12} and agree with that of \cite{qiu11}.
In addition, as discussed in Section 4.6.1, we might underestimate
the size of the infalling region, and thus overestimate the characteristic
excitation temperature of the $^{13}$CO gas in this region. However, since
in equation (3) the column density $N_{\rm H}$ is proportional to
excitation temperature, we expect that the different approach (larger
infalling region, low characteristic T$_{\rm ex}$) would
not significantly change the derived infall rate.

The infall rate can also be estimated from the
HCN(1-0) results despite larger uncertainty in the
the line profile fitting, and the poor determination of the abundance.
On the basis of \cite{rohlf04}, the column density of HCN can
be derived by:
\begin{eqnarray}
N({\rm HCN}) &=& 5.41\times10^{11}{\rm cm}^{-2}\frac{T_{\rm ex}\int \tau_v{\rm d}v}
{1-{\rm exp}(-4.25/T_{\rm ex})}
\end{eqnarray}
With $T_{\rm ex}=200$ K, the two redshifted absorption HCN(1-0)
components at $-54.0$ and $-57.0$ km s$^{-1}$ give
HCN column densities of $N_{\rm HCN}=5\times10^{16}$ and
$N_{\rm HCN}=1\times10^{16}$ cm$^{-2}$, respectively.
A recent study
reported a H$^{13}$CN abundance of $\sim$2$\times$10$^{-10}$ in
active star formation regions \citep{zinc09}, so we estimate
a total H$_2$ column density $N(\rm H_2)=5\times10^{24}$ cm$^{-2}$
with ${\rm [H^{13}CN]/[HCN]}=1/60$,
which corresponds to an infall rate of 1$\times$10$^{-2}$ $\rm M_\odot$ yr$^{-1}$,
triple of the value derived from $^{13}$CO(2-1).
Although the infall rates derived from
$^{13}$CO (2-1) and HCN(1-0) are relatively high, they are still in the
typical range of 10$^{-4}$ to 10$^{-2}$ M$_\sun$ yr$^{-1}$ for embedded
high-mass cores \cite[][and the references therein]{wu09}.

We can compare the derived infall rates with an ideal free-fall case.
With an enclosed mass of 30 M$_\sun$
(including the masses of both gas and the central star)
at a radius $r=$0\farcs7, the free-fall infall velocity $V_{\rm in, ff}$
is about 5.5 km s$^{-1}$, corresponding to a free-fall infall
rate of 5$\times$10$^{-3}$ M$_\sun$ yr$^{-1}$ from the $^{13}$CO(2-1) results.
This value agrees with 3$\times$10$^{-3}$ M$_\sun$ yr$^{-1}$, calculated
from the velocity displacements of the redshifted
$^{13}$CO (2-1) absorption features, indicating that gas is falling into IRS\,1
rapidly.

In addition to the redshifted absorption velocity components, we see
a blueshifted absorption at $V_{\rm lsr}=-63.8$ km s$^{-1}$ in $^{13}$CO(2-1)
or $V_{\rm lsr}=-63.7$ km s$^{-1}$ in CO(2-1) and HCN(1-0).
Since the optical depths of the blueshifted absorption are significantly lower
than the redshifted ones, and no other observations have suggested a cloud
component in NGC\,7538 region at this radial velocity which is only seen toward
IRS\,1, we believe that the absorption line components at $-63.7$ or $-63.8$
km s$^{-1}$ arise from the blueshifted outflow gas between IRS\,1 and the observer.

\subsection{The Nature of Hot Molecular Clumps \& Dynamics in NGC\,7538 IRS\,1}

The kinematics observed from
IRS\,1 SW and NE in the central 2$\arcsec$, from OCS(19-18),
CH$_3$CN(12-11) and $^{13}$CO(2-1) show a velocity gradient
in P.A.$\sim$43$\arcdeg$.
The three molecular species observed in this paper are in good
agreement with the results from the $^{13}$CO(1-0) emission
\citep{scov86} and $^{15}$NH$_3$ masers \citep{gaum91}
observations.

The observed high excitation temperatures and
velocities of the hot molecular components in the
circumstellar region of IRS 1 provide clues for understanding
the IRS\,1 system. Infrared radiation from IRS 1 supplies the
primary heating in the central 1-2\arcsec, ($\sim$5,000 AU).
The ionized emission observed with the VLA in the inner
$<2\arcsec$\citep{camp84,gaum95,sand09} is suggested to be a
hot ionized wind with a velocity up to 300 km s$^{-1}$ which
is partially confined and collimated by the molecular
material surrounding IRS\,1 \citep{gaum95}. The hot
molecular components (see Figure 6a and 9a) are located NE of
the ionized core, and SW of the southern lobe of the ionized
outflow. These two locations are close to the path of the
ionized outflow, suggesting that the molecular gas may be heated
by the interaction between the ionized outflow and accreting
gas.

A large accretion flow with a rate of 3-10$\times10^{-3}$
M$_\odot$ y$^{-1}$ continues to accumulate matter into the
circumstellar region around IRS 1 (see Section 5.2).
Confronting the overwhelming pressure from the accretion
flow, the HC \ion{H}{2} region surrounding the O star must
be highly confined, and difficult to expand to form a
larger size \ion{H}{2} region. The excess angular
momentum in the infalling gas must be transported out.
The outflow provides a mechanism for transferring the angular
momentum of the infalling gas.

The large-scale CO(2-1), $^{13}$CO(1-0) and HCO$^+$(1-0) images from
CARMA, SMA and single-dish observations suggest a
bipolar outflow in a north-south direction with an opening
angle greater than $\sim$90\arcdeg \citep{sand12}. The
interaction between this wide-angle molecular outflow and
the envelope of ambient molecular gas around IRS\,1 might
excite the observed hot molecular components, redshifted to the NE,
and blueshifted to the SW. The density of the environment is
much higher to the south of IRS\,1; thus the molecular
emission from IRS\,1 SW is stronger, and the ionized outflow
in this direction is highly confined.

The redshifted radial velocity of IRS\,1 NE could also
be due to rotation of the wide-angle outflow around its
axis. An ionized outflow which rotates around its axis could
entrain the adjacent molecular gas, and consequently, the
impact of the ionized outflow on the circumstellar gas could
produce highly-excited molecular species at the positions of IRS\,1 NE and SW.

The ambient density is higher to the south and SW of IRS\,1, so
the entrainment of the ionized jet is not significant.
To the north of IRS\,1, the density is lower,
and the ionized outflow can penetrate
and produce significant redshifted movement of the entrained
gas NE of IRS\,1. We can roughly estimate the angular
momentum of the NE component to be 3.6$\times$10$^{17}$ M$_\sun$
m$^2$ s$^{-1}$ by assuming a mass $M_{\rm NE}=1$ M$_\sun$ (see
Section 4.5), rotation velocity $v_{\rm NE}=8$ km s$^{-1}$, and
radius $r_{\rm NE}=300$ AU ($\Delta \alpha=0\farcs11$, see Table
4). The angular momentum input to IRS\,1 is
about 5.7$\times$10$^{15}$ M$_\sun$ m$^2$ s$^{-1}$ yr$^{-1}$ with
an infall rate of 5$\times$10$^{-3}$ M$_\sun$ yr$^{-1}$, a radius of
1900 AU ($\sim$0\farcs7) and an enclosed mass of 30 M$_\sun$.
Therefore, it is plausible that the ionized outflow could cause the
observed redshifted movement of the component NE in the lifetime of
the accretion process.

\section{Conclusion}

Using sub-arcsec resolution observations at 1.3 and 3.4 mm
with SMA and CARMA, we analyzed the molecular lines and
continuum emission from NGC\,7538\,IRS\,1 to investigate the
dynamic processes related to the star-forming activities in
this region. With an image convolved to 0\farcs1 from
CARMA B-array observation at 1.3 mm, the primary continuum
source IRS 1, an HC \ion{H}{2} region, was resolved into two
components, an unresolved compact core with a linear size
$<270$ AU and a north-south extended lobe of the ionized
outflow in agreement with the predicted
extrapolation of a power-law dependence made by \cite{sand09}.
Both the SMA and CARMA show
extended dust continuum emission to the SW of IRS\,1, which might
indicate a reservoir of molecular gas.

The molecular lines OCS(19-18), CH$_3$CN(12-11)
and $^{13}$CO(2-1), show that dense molecular
gas is primarily distributed southwest to northeast
across IRS\,1 with an angular size of
$\sim$2\arcsec$\times$1\arcsec.
The locations and velocities of the molecular clumps
suggest that they might be heated by
interaction between the ionized outflow from IRS\,1 and the
adjacent molecular gas in a rotating wide-angle
outflow in a N-S direction from IRS\,1.

From high-resolution (0\farcs8$\times$0\farcs6) observations of
$^{13}$CO(2-1), CO(2-1) and HCN(1-0), we fitted multiple
Gaussian velocity components to the spectra toward the
continuum sources in IRS 1. We found that the significant molecular
lines in absorption are redshifted with
respect to the systemic velocity toward IRS 1, suggesting
that a large amount of material surrounding the region
is infalling toward IRS\,1. An infall rate of
$\sim$3-10$\times$10$^{-3}$ M$_\sun$ y$^{-1}$ was inferred
from $^{13}$CO(2-1) and HCN(1-0) lines.

\acknowledgments
LZ is supported by National Basic Research Program of China
(973 program) No. 2012CB821802. He was a SAO predoctoral
fellow, and a part of the work in this paper was carried out
during the course of his Ph.D. research. JHZ is grateful to
the National Astronomical Observatories of China for hosting
his visit during the course of writing this research paper.
We are grateful to Miller Goss for his helpful comments and
suggestions. Support for CARMA construction was derived from
the states of California, Illinois, and Maryland, the Gordon
and Betty Moore Foundation, the Kenneth T. and Eileen L.
Norris Foundation, the Associates of the California
Institute of Technology, and the National Science Foundation.
Ongoing CARMA development and operations are supported by
the National Science Foundation under a cooperative agreement,
and by the CARMA partner universities. The Very Large Array
(VLA) is operated by the National Radio Astronomy Observatory
(NRAO). The NRAO is a facility of the National Science
Foundation operated under cooperative agreement by
Associated Universities, Inc.

\clearpage
\begin{deluxetable}{lcccccc}
\tablenum{1}
\tablecaption{Log of Observation, uv data and images}
\tablehead{
\multicolumn{7}{c}{VISIBILITY DATA SETS}\\
\colhead{Obs. Date}&
\multicolumn{2}{c}{Obs. pointing}&
\colhead{Array Conf.}&
\colhead{} &
\colhead{Calibrators}&
\colhead{}  \\
\colhead{} &
\colhead{$\alpha_{\rm J2000}$}&
\colhead{$\delta_{\rm J2000}$}&
\colhead{} &
\colhead{phase} &
\colhead{bpass} &
\colhead{flux~~~~~~~~~~}
}
\startdata
\sidehead{SMA at 1.3 mm}
2008-05-05 &23:13:45.362&61:28:10.49& VEX &J2202+422&3C454.3&Vesta \\
2005-09-11 &23:13:45.300&61:28:10.00& EXT &J0102+584&3C454.3&Ceres \\
           &            &           &     &J2202+422&Neptune&      \\
\sidehead{CARMA at 1.3  mm}
2010-01-04 &23:13:45.600&61:28:11.00& B-array &J0102+584&3C454.3&Uranus, MWC349 \\
2010-01-05 &23:13:45.600&61:28:11.00& B-array &J0102+584&3C454.3&Neptune \\
2010-11-10 &23:13:45.600&61:28:11.00& C-array &J0102+584&3C454.3&Uranus \\
2009-11-23 &23:13:45.600&61:28:11.00& C-array &J0102+584&3C84   &Uranus \\
2009-11-27 &23:13:45.600&61:28:11.00& C-array &J0102+584&3C454.3&MWC349 \\
2010-11-29 &23:13:45.600&61:28:11.00& C-array &J0102+584&3C454.3&Neptune \\
2007-03-15 &23:13:45.600&61:28:11.00& C-array &J0102+584&3C454.3&MWC349 \\
2010-08-24 &23:13:45.600&61:28:11.00& D-array &J0102+584&3C454.3&Uranus \\
2009-08-07 &23:13:45.600&61:28:11.00& D-array &J0102+584&3C454.3&Neptune \\
2010-08-27 &23:13:45.600&61:28:11.00& D-array &J0102+584&3C454.3&Neptune \\
2010-08-27 &23:13:45.600&61:28:11.00& D-array &J0102+584&3C454.3&Neptune \\
2010-09-08 &23:13:45.600&61:28:11.00& D-array &J0102+584&3C454.3&Uranus \\
2010-07-05 &23:13:45.600&61:28:11.00& E-array &J0102+584&3C454.3&MWC349 \\
2010-07-07 &23:13:45.600&61:28:11.00& E-array &J0102+584&3C454.3&MWC349 \\
2010-07-08 &23:13:45.600&61:28:11.00& E-array &J0102+584&3C454.3&URANUS \\
\sidehead{CARMA at 3.4 mm}
2010-12-29 &23:13:45.298&61:28:10.00& B-array &J0102+584&3C454.3&Uranus \\
\\
\tableline \\
\multicolumn{7}{c}{CONTINUUM \& LINE IMAGES} \\
{Mode     }& FWHM beam  & rms       & BW   &$\delta V$&$V_{\rm LSR}$-ranges&uv
datasets\\
    &\arcsec$\times$\arcsec(\arcdeg)& mJy b$^{-1}$ (ch$^{-1}$)&GHz& km s$^{-1}$ &
    km s$^{-1}$ \\
\tableline
\sidehead{SMA}
Continuum 1.3 mm &0.8$\times$0.6 (--45)& 4& 4 &\nodata  &\nodata&VEX + EXT\\
Line OCS(19-18) &0.8$\times$0.6 (--45)&60&\nodata&0.5     &--85 to --35&VEX + EXT\\
Lines CH$_3$CN(12-11) &0.8$\times$0.6 (--45)&60&\nodata&0.5&--85 to --35&VEX + EXT\\
Line $^{13}$CO (2-1) &0.8$\times$0.6 (--45)&50&\nodata&0.5&--110 to --10&VEX + EXT \\
Line $^{12}$CO (2-1) &0.8$\times$0.6 (--45)&60&\nodata&1&--160 to +40&VEX + EXT \\
\sidehead{CARMA}
Continuum 1.3 mm &0.78$\times$0.64 (--84)& 2.5&2-3  &\nodata  &\nodata&B+C+D+E-array\\
Continuum 1.3 mm &0.34$\times$0.27 (--83)& 10& 2  &\nodata  &\nodata&B-array\\
Continuum 3.4 mm &0.8$\times$0.7 (--68)& 7& 0.064  &\nodata  &\nodata&B-array\\
Line HCN(1-0)    &0.8$\times$0.7 (--68)& 33&\nodata& 1.65 &--220 to --11&B-array\\
\enddata
\end{deluxetable}

\begin{deluxetable}{lcccc}
\tablenum{2}
\tablecaption{Results from continuum observations at 1.3 \& 3.4 mm}
\tablehead{
\colhead{Source}&
\colhead{$S_{\rm p}$}&
\colhead{$S_{\rm t}$}&
\colhead{$\Delta \alpha$, $\Delta \delta$}&
\colhead{$\Delta \theta_{\rm M}\times\Delta \theta_{\rm m}$ (P.A.)}\\
\colhead{} &
\colhead{(Jy beam$^{-1}$)}&
\colhead{(Jy)}&
\colhead{(\arcsec, \arcsec)}&
\colhead{(\arcsec$\times$\arcsec (\arcdeg))}
}
\startdata
\multicolumn{5}{c}{SMA Results} \\
\sidehead{SMA 1.3 mm}
IRS\,1&3.3$\pm$0.3 &3.5$\pm$0.4 & 0.0$\pm$0.1,  0.0$\pm$0.1 &unresolv\\
SW Extenstion&$0.12\pm$0.02 &$0.25\pm$0.04& --0.5$\pm$0.1, --0.9$\pm$0.1 &1.0$\times$0.6 (22)\\
\\
\tableline\\
\sidehead{CARMA 1.3 mm}
IRS\,1-Peak&1.0$\pm$0.1 &1.0$\pm$0.1 & --0.08$\pm$0.02,  --0.07$\pm$0.02 & unresolv \\
IRS\,1-Lobe&0.22$\pm$0.02 &2.8$\pm$0.3 & --0.08$\pm$0.01,  --0.09$\pm$0.01 &0.42$\times$0.28 (--2)\\
IRS\,1-total&\nodata      &3.8$\pm$0.4  &\nodata                           &\nodata    \\
\sidehead{CARMA 3.4 mm}
IRS\,1&0.93$\pm$0.10 &0.93$\pm$0.10 & $0.0\pm$0.1,  $0.0\pm$0.1 &unresolv\\
\enddata
\tablecomments{The reference position is R.A. (J2000) = 23$^{\rm h}$13$^{\rm m}$45.$^{\rm s}$37,
Dec. (J2000) = 61$\arcdeg$28$\arcmin$10\farcs43}
\end{deluxetable}

\begin{deluxetable}{ccccc}
\tablenum{3}
\tablecaption{Fitting of the high transition lines at 0.86 mm}
\tablehead{
\colhead{Transition}&
\colhead{$\nu$}&
\colhead{$\Delta S_{\rm L}$}&
\colhead{$V_{\rm LSR}$}&
\colhead{$\Delta$$V_{\rm FWHM}$}\\
\colhead{~} &
\colhead{(GHz)} &
\colhead{(Jy beam$^{-1}$)}&
\colhead{(km s$^{-1}$)}&
\colhead{(km s$^{-1}$)}
}
\startdata
CH$_3$OH(12,1,11-12,0,12)        & 336.865 & 27.8$\pm$0.4 & --58.7$\pm$0.1 & 4.8$\pm$0.2 \\
$^{13}$CH$_3$OH(12,1,11-12,0,12) & 335.560 & ~6.2$\pm$0.2 & --58.5$\pm$0.1 & 3.1$\pm$0.2 \\
$^{13}$CH$_3$OH(14,1,13-14,0,14) & 347.188 & ~4.9$\pm$0.2 & --58.6$\pm$0.1 & 3.4$\pm$0.2 \\
\enddata
\end{deluxetable}

\begin{deluxetable}{lcccccccccc}
\tablenum{4}
\tabletypesize{\small}
\tablecaption{Properties of Hot Molecular Components in IRS\,1}
\tablehead{
Transition            &
$\nu$                 &
$\Delta\alpha,\Delta\delta $       &
$\Delta S_{\rm L}$ &
$V_{\rm LSR}$     &
$\Delta V_{\rm FWHM}$              &
$\displaystyle \int T_{\rm B} {\rm dv} {\rm (peak)}$ &
$N^{\rm obs}_{JK}$  &
$N^{\rm corr}_{JK}$  &
$g_{JK}$  &
$E_{JK}/k$ \\
      &
(GHz) &
(\arcsec)   &
(Jy b$^{-1}$) &
(km s$^{-1}$) &
(km s$^{-1}$) &
(K km s$^{-1}$) &
\multicolumn{2}{c}{(10$^{13}$ cm$^{-2}$)} &
              &
(K)
}
\startdata \\
\multicolumn{11}{c}{IRS 1 SW}\\
\sidehead{CH$_3$CN}
12$_0$-11$_0$ & 220.747 &
--0.16$\pm$0.04, --0.44$\pm$0.03& 1.4$\pm$0.1 & --59.3$\pm$0.1 & 4.4$\pm$0.2&282$\pm$22
&2.9$\pm$0.2 &\nodata&100&68 \\
12$_1$-11$_1$ & 220.743 &
--0.15$\pm$0.06, --0.45$\pm$0.05& 1.4$\pm$0.1 & --59.2$\pm$0.1 & 3.9$\pm$0.3&258$\pm$21 
&2.7$\pm$0.2&\nodata&100&76 \\
 12$_2$-11$_2$ & 220.730 &
--0.11$\pm$0.04, --0.45$\pm$0.03& 1.4$\pm$0.1 & --59.6$\pm$0.1 & 4.2$\pm$0.2&260$\pm$24 
&2.8$\pm$0.3&\nodata&100&97 \\
 12$_3$-11$_3$ & 220.709 &
--0.21$\pm$0.04, --0.48$\pm$0.03& 1.4$\pm$0.1 & --59.6$\pm$0.1 & 4.3$\pm$0.1&282$\pm$18 
&2.8$\pm$0.2&\nodata&200&133 \\
 12$_4$-11$_4$ & 220.679 &
--0.07$\pm$0.03, --0.43$\pm$0.03& 1.2$\pm$0.1 & --59.5$\pm$0.1 & 4.2$\pm$0.2&249$\pm$23 
&2.8$\pm$0.3&3.3$\pm$0.3&100&183 \\
 12$_5$-11$_5$ & 220.641 &
--0.04$\pm$0.04, --0.41$\pm$0.03& 0.9$\pm$0.1 & --59.5$\pm$0.1 & 4.6$\pm$0.3&201$\pm$26 
&2.5$\pm$0.3&2.8$\pm$0.4&100&247 \\
 12$_6$-11$_6$ & 220.594 &
--0.01$\pm$0.04, --0.30$\pm$0.03& 0.9$\pm$0.1 & --59.5$\pm$0.1 & 4.5$\pm$0.2&205$\pm$20 
&2.8$\pm$0.3&3.2$\pm$0.3&200&326 \\
 12$_7$-11$_7$ & 220.539 &
+0.08$\pm$0.09, --0.25$\pm$0.08 & 0.5$\pm$0.1 & --59.5$\pm$0.2 & 4.1$\pm$0.4&97$\pm$23  
&1.5$\pm$0.4&1.5$\pm$0.4&100&419 \\
 12$_8$-11$_8$ & 220.476 &
+0.06$\pm$0.10, --0.45$\pm$0.12& 0.3$\pm$0.1 & --59.3$\pm$0.2 & 3.4$\pm$0.4&54$\pm$12   
&1.0$\pm$0.2&1.0$\pm$0.2&100&526 \\
($K$=2-5)     &           &
--0.11$\pm$0.03, --0.44$\pm$0.02& 1.2$\pm$0.1 & --59.5$\pm$0.1 & 4.3$\pm$0.1&250$\pm$16 
&\nodata&\nodata&\nodata&\nodata \\
\sidehead{OCS }
19-18         & 231.061 &
--0.13$\pm$0.03, --0.41$\pm$0.03& 1.3$\pm$0.1 & --59.5$\pm$0.1 & 4.4$\pm$0.1&267$\pm$15 & \nodata&\nodata&\nodata&\nodata \\
\tableline
\multicolumn{11}{c}{} \\
\multicolumn{11}{c}{IRS\,1 NE} \\
\sidehead{CH$_3$CN  }
12$_0$-11$_0$ & 220.747 &
+0.39$\pm$0.15, +0.26$\pm$0.25& 0.6$\pm$0.1 & --52.0$\pm$0.2 & 5.4$\pm$0.5&164$\pm$22 
&1.4$\pm$0.2&1.5$\pm$0.2&100&69 \\
12$_1$-11$_1$ & 220.743 &
+0.21$\pm$0.03, +0.33$\pm$0.03& 0.5$\pm$0.1 & --51.4$\pm$0.4 & 5.4$\pm$0.5&138$\pm$24 
&1.4$\pm$0.2&1.5$\pm$0.3&100&76 \\
12$_2$-11$_2$ & 220.730 &
+0.23$\pm$0.04, +0.36$\pm$0.06& 0.4$\pm$0.1 & --51.4$\pm$0.3 & 6.0$\pm$0.7&109$\pm$23 
&1.2$\pm$0.2&1.2$\pm$0.3&100&97 \\
12$_3$-11$_3$ & 220.709 &
+0.26$\pm$0.06, +0.28$\pm$0.07& 0.3$\pm$0.1 & --51.0$\pm$0.3 & 5.7$\pm$0.6&96$\pm$14  
&1.1$\pm$0.2&\nodata&200&133 \\
12$_4$-11$_4$ & 220.679 &
+0.14$\pm$0.15, +0.35$\pm$0.16& 0.3$\pm$0.1 & --51.3$\pm$0.3 & 4.4$\pm$0.7&61$\pm$18  
&0.7$\pm$0.2&0.7$\pm$0.2&100&183 \\
12$_5$-11$_5$ & 220.641 &
+0.31$\pm$0.06, +0.12$\pm$0.08& 0.3$\pm$0.1 & --52.0$\pm$0.5 & 4.1$\pm$0.6&49$\pm$15  
&0.6$\pm$0.2&0.6$\pm$0.2&100&247 \\
12$_6$-11$_6$ & 220.594 &
+0.24$\pm$0.06, +0.27$\pm$0.08& 0.4$\pm$0.1 & --50.8$\pm$0.2 & 3.4$\pm$0.2&60$\pm$10  
&0.8$\pm$0.1&0.9$\pm$0.2&200&326 \\
12$_7$-11$_7$ & 220.539 &
+0.42$\pm$0.15, +0.06$\pm$0.16& 0.2$\pm$0.1 & --51.1$\pm$0.3 & 2.8$\pm$0.5&26$\pm$11  
&0.4$\pm$0.2&0.4$\pm$0.2&100&419 \\
 ($K$=2-5)     &           &
+0.23$\pm$0.04, +0.33$\pm$0.04& 0.3$\pm$0.1 & --51.4$\pm$0.2 & 6.8$\pm$0.5 &103$\pm$14 &\nodata 
&\nodata&\nodata&\nodata \\
\sidehead{OCS}
19-18         & 231.061 &
+0.25$\pm$0.03, +0.35$\pm$0.02& 0.5$\pm$0.1 & --51.5$\pm$0.3 & 7.2$\pm$0.6 &180$\pm$30 &\nodata &\nodata&\nodata &\nodata\\
\enddata
\tablecomments{The reference position is R.A. (J2000) = 23$^{\rm h}$13$^{\rm m}$45.$^{\rm s}$37,
Dec. (J2000) = 61$\arcdeg$28$\arcmin$10\farcs43}
\end{deluxetable}

\begin{deluxetable}{lcrrrrrr}
\tablenum{5}
\tablecaption{$^{13}$CO(2-1), CO(2-1) and HCN(1-0) toward IRS\,1}
\tablehead{
\colhead{Source}&
\colhead{$\Delta \alpha$, $\Delta \delta$}&
\colhead{$\Delta S_{\rm L}$ }&
\colhead{$V_{\rm LSR}$} &
\colhead{$\Delta V_{\rm FWHM}$}&
\colhead{$\tau_{\rm L}$}&
\colhead{$N_{\rm ^{13}CO}$(200)$^a$} &
\colhead{$N_{\rm ^{13}CO}$(75)$^b$}  \\
\\
\colhead{} &
\colhead{\arcsec, \arcsec} &
\colhead{(Jy)}&
\colhead{(km s$^{-1}$)}&
\colhead{(km s$^{-1}$)}&
\colhead{}&
\colhead{(10$^{15}$ cm$^{-2}$)} &
\colhead{(10$^{15}$ cm$^{-2}$)}
}
\startdata \\
\multicolumn{8}{c}{$^{13}$CO(2-1) with a beam area 0\farcs8$\times$0\farcs6} \\
\\
\tableline \\
IRS\,1  &+0.0, +0.0 &$-0.5\pm$0.1 & --63.8$\pm0.3$ & 1.3$\pm$0.2 &0.2$\pm$0.0 &110 &40 \\
        &           &$0.5\pm$0.1 & --63.0$\pm0.3$ & 5.5$\pm$0.4 &0.2$\pm$0.0 &490 &180 \\
        &           &$-2.6\pm$0.1 & --59.0$\pm0.3$ & 1.5$\pm$0.2 &1.6$\pm$0.1 &1300 &470 \\
        &           &$-2.3\pm$0.1 & --57.0$\pm0.3$ & 2.5$\pm$0.2 &1.2$\pm$0.1 &1600 &610  \\
        &           &$-2.0\pm$0.1 & --54.5$\pm0.3$ & 2.5$\pm$0.2 &1.0$\pm$0.1 &1500 &560  \\
        &           &$\phantom{+}0.5\pm$0.1 & --48.0$\pm0.3$ & 2.5$\pm$0.2 &0.2$\pm$0.0 &240 &90 \\
\\
SW Extension &$-$0.5, $-$0.9 &0.6$\pm$0.1 & --60.1$\pm$0.1 & 1.8$\pm$0.1 &\nodata&\nodata&\nodata \\
            &               &0.2$\pm$0.0 & --63.2$\pm$0.1 & 2.1$\pm$0.0 &\nodata&\nodata&\nodata \\
\\
\tableline \\
\multicolumn{8}{c}{CO(2-1) with a beam area 0\farcs8$\times$0\farcs6} \\
\\
\tableline \\
IRS\,1 &+0.0, +0.0 &--1.7$\pm$0.2 & --63.7$\pm0.3$ & 7.2$\pm$0.5 &0.7$\pm$0.2&\nodata &\nodata \\
        &          &--2.5$\pm$0.4 & --59.0$\pm0.3$ & 2.0$\pm$0.2 &1.5$_{-0.5}^{+0.7}$ &\nodata &\nodata \\
        &          &--2.7$\pm$0.4 & --57.0$\pm0.3$ & 2.8$\pm$0.2 &2.0$_{-0.8}^{+1.0}$ &\nodata &\nodata \\
        &          &--3.2$\pm$0.3 & --54.0$\pm0.3$ & 3.2$\pm$0.3 &4.1$_{-1.9}^{>5.0}$ &\nodata &\nodata \\
        &          &1.3$\pm$0.2 & --48.2$\pm0.3$ & 3.5$\pm$0.2 &\nodata &\nodata &\nodata \\
\tableline \\
\multicolumn{8}{c}{HCN(1-0) with a beam area 0\farcs8$\times$0\farcs7} \\
\\
\tableline \\
IRS\,1  &+0.0, +0.0 &--0.7$\pm$0.1 &\nodata & 7.1\phantom{$\pm$0.6}&\nodata &\nodata &\nodata \\
        &           &--0.8$\pm$0.1 & --63.7$\pm0.3$ & 7.1$\pm$0.6&1.3$_{-0.3}^{+0.5}$ &\nodata &\nodata \\
        &           &--0.3\phantom{$\pm$0.1} &\nodata & 7.1\phantom{$\pm$0.6}&\nodata &\nodata &\nodata \vspace{.15cm}\\
        &           &--0.1\phantom{$\pm$0.1} &\nodata & 2.0\phantom{$\pm$0.1} &\nodata &\nodata &\nodata \\
        &           &--0.7$\pm$0.1 &--59.0$\pm0.3$ & 2.0$\pm$0.1 &1.1$_{-0.2}^{+0.4}$ &\nodata &\nodata \\
        &           &--0.2\phantom{$\pm$0.1} &\nodata & 2.0\phantom{$\pm$0.1} &\nodata &\nodata &\nodata \vspace{.15cm}\\
        &           &--0.1\phantom{$\pm$0.1} &\nodata & 2.8\phantom{$\pm$0.1} &\nodata &\nodata &\nodata \\
        &           &--0.6$\pm$0.1 & --57.0$\pm0.3$ & 2.8$\pm$0.2 &0.8$_{-0.2}^{+0.2}$ &\nodata &\nodata \\
        &           &--0.2\phantom{$\pm$0.1} &\nodata & 2.8\phantom{$\pm$0.1} &\nodata &\nodata &\nodata \vspace{.15cm}\\
        &           &--0.4\phantom{$\pm$0.1} &\nodata & 5.2\phantom{$\pm$0.1} &\nodata &\nodata &\nodata \\
        &           &--1.1$\pm$0.1 & --54.0$\pm$0.3 & 5.2$\pm$0.3 &1.8$_{-0.4}^{+0.7}$ &\nodata &\nodata \\
        &           &--1.0$\pm$0.1 &\nodata &5.2\phantom{$\pm$0.1} &\nodata &\nodata &\nodata \\
\enddata
\tablecomments{
\noindent The reference position is R.A. (J2000) = 23$^{\rm h}$13$^{\rm m}$45.$^{\rm s}$37,
Dec. (J2000) = 61$\arcdeg$28$\arcmin$10\farcs43 \\
$^a$ with $T_{\rm ex}=200$ K\\
$^b$ with $T_{\rm ex}=75$ K}
\end{deluxetable}
\end{document}